\documentclass[twocolumn]{aastex62}
\usepackage{graphicx}   
\usepackage{amsmath}    
\usepackage{amssymb}    
\usepackage{threeparttable}
\usepackage{ulem}

\graphicspath{{./}{figures/}}

\received{June 13, 2018}
\revised{}
\accepted{July 9, 2018}

\shorttitle{Why Are Some GRBs Hosted by Oxygen-rich Galaxies?}
\shortauthors{Hashimoto et al.}

\begin{document}

\title{Why Are Some Gamma-Ray Bursts Hosted by Oxygen-rich Galaxies?}

\correspondingauthor{Tetsuya Hashimoto}
\email{tetsuya@phys.nthu.edu.tw}

\author{Tetsuya Hashimoto}
\affiliation{Institute of Astronomy, National Tsing Hua University \\
101, Section 2. Kuang-Fu Road, Hsinchu, 30013, Taiwan}

\author{Ravi Chaudhary}
\affiliation{Department of Physics, Indian Institute of Technology Delhi \\
Hauz Khas, New Delhi, 110016, India}

\author{Kouji Ohta}
\affiliation{Department of Astronomy, Kyoto University \\
Kitashirakawa Oiwake-cho, Sakyo-ku, Kyoto, Kyoto, 606-8588, Japan}

\author{Tomotsugu Goto}
\affiliation{Institute of Astronomy, National Tsing Hua University \\
101, Section 2. Kuang-Fu Road, Hsinchu, 30013, Taiwan}

\author{Francois Hammer}
\affiliation{Laboratoire Galaxies {\'E}toiles Physique et Instrumentation, Observatoire de Paris-Meudon\\
5 place Jules Janssen, F-92195 Meudon Cedex, France}

\author{Albert K. H. Kong}
\affiliation{Institute of Astronomy, National Tsing Hua University \\
101, Section 2. Kuang-Fu Road, Hsinchu, 30013, Taiwan}

\author{Ken'ichi Nomoto}
\affiliation{Kavli Institute for the Physics and Mathematics of the Universe (WPI), The University of Tokyo \\
Kashiwa, Chiba 277-8583, Japan}

\author{Jirong Mao}
\affiliation{Yunnan Observatories, Chinese Academy of Sciences\\ 
650011 Kunming, Yunnan Province, China}



\begin{abstract}
Theoretically long gamma-ray bursts (GRBs) are expected to happen in low-metallicity environments, because in a single massive star scenario, low iron abundance prevents loss of angular momentum through stellar wind, resulting in ultra-relativistic jets and the burst.
In this sense, not just a simple metallicity measurement but also low iron abundance ([Fe/H]$\lesssim$-1.0) is essentially important.
  
Observationally, however, oxygen abundance has been measured more often due to stronger emission. 
In terms of oxygen abundance, some GRBs have been reported to be hosted by high-metallicity star-forming galaxies, in tension with theoretical predictions.

Here we compare iron and oxygen abundances for the first time for GRB host galaxies (GRB 980425 and 080517) based on the emission-line diagnostics. 
The estimated total iron abundances, including iron in both gas and dust, are well below the solar value. 
The total iron abundances can be explained by the typical value of theoretical predictions ([Fe/H]$\lesssim$-1.0), despite high oxygen abundance in one of them.
According to our iron abundance measurements, the single massive star scenario still survives even if the oxygen abundance of the host is very high, such as the solar value. 
Relying only on oxygen abundance could mislead us on the origin of the GRBs.

The measured oxygen-to-iron ratios, [O/Fe], can be comparable to the highest values among the iron-measured galaxies in the Sloan Digital Sky Survey. 
Possible theoretical explanations of such high [O/Fe] include the young age of the hosts, top-heavy initial mass function, and fallback mechanism of the iron element in supernova explosions.

\end{abstract}



\section{Introduction} 
\label{introduction}
Observationally, there is a population of gamma-ray bursts (GRBs) that shows \lq long\rq\ durations ($>2$s) of bursts \citep{Kouveliotou1993}.
It is widely accepted that \lq long\rq\ GRB (hereafter just referred to as GRB) emission is emitted from ultra-relativistic jets that are launched at the core collapse of a single massive star. 
This scenario requires a large angular momentum in the collapsing core to launch such jets \citep[e.g.,][]{Woosley2006}. 
In this scenario, the angular momentum is removed by the stellar wind driven by radiation pressure that depends on the iron abundance \citep{Vink2005}.
Thus, low iron abundance ([Fe/H]$\lesssim$ -1.0) is favored in this single massive star scenario, because the progenitor with low iron abundance does not lose its angular momentum by the mass loss during its evolution \citep[e.g.,][]{Yoon2006}. 
Observationally, oxygen abundance has been used as an indicator of iron abundance rather than iron abundance itself, since oxygen has strong emission lines in the rest-frame optical wavelength.
Indeed, optical emission-line diagnostics show that many of the GRB hosts have the subsolar oxygen abundances \citep[e.g.,][]{Stanek2006,Modjaz2008,Levesque2010a,Levesque2010b}.



However, recent observational efforts to investigate unbiased host samples indicate that GRBs can occur in dusty massive star-forming galaxies \citep[e.g.,][]{Hashimoto2010,Perley2016a,Perley2016b}, implying high oxygen abundance.
In fact, some of these host galaxies have been confirmed to show high oxygen abundance approaching the solar value based on the emission-line diagnostics \citep[e.g.,][]{Graham2013,Hashimoto2015,Kruhler2015,Stanway2015}.
The possible high-metallicity environment of GRBs has also been reported from the absorption system in the optical afterglow \citep[e.g.,][]{Savaglio2012}.

Why is the oxygen abundance high despite the theoretical expectation?
There could be three possible explanations for such a GRB environment with high oxygen abundance. 
One is the metallicity dispersion of star-forming region within a host galaxy.
The spatial resolutions of abundance measurements at the GRB positions (typically kpc scale) are still not enough to resolve each star-forming regions \citep{Niino2015}, except for particularly near GRB host galaxies such as 980425 \citep{Hammer2006}. 
In this sense, observed high oxygen abundances (regardless of averaged over the whole host galaxy or at the GRB position) do not exclude the hypothesis that the GRBs are born in environments with low oxygen abundances.

Secondly, there may be another channel for massive stars to become GRBs, other than the single massive star scenario.
For instance, a binary star merger explosion scenario may be important to GRBs that are hosted by galaxies with high oxygen abundance.
In this scenario, the merging stars take in their orbital angular momentum during the merging process. 
The sufficient angular momentum required for a GRB explosion is maintained even in the high-metallicity environment \citep[e.g.,][]{Nomoto1995,Fryer1999}.

The third possible explanation is the overabundance of oxygen, i.e., a high value of [O/Fe] in GRB host galaxies. 
In such a case, the oxygen abundance can be as high as the solar value, even if the iron abundance is very low, as expected from the single massive star scenario mentioned above.
It is well known that the iron is mainly produced by Type Ia supernovae \citep[SNe Ia;][]{Tinsley1979}, whereas $\alpha$ elements such as oxygen, neon and sulfur are produced by Type II supernovae (SNe II). 
The time scale of SNe Ia ($\sim 10^{9}$ yr) is much longer than that of SNe II ($\sim 10^{7}$ yr). 
Therefore it is possible that a very young galaxy has no (or less) experience with chemical enrichment from SN Ia but has contribution from SN II, which results in a high value of [O/Fe] \citep[e.g.,][]{McWilliam1997}.
In fact, the overabundance of oxygen has been confirmed for galaxies with strong emission lines \citep{Izotov2006}, which are probably dominated by very young stellar populations undergoing active star formation.
Such may be the case with GRB host galaxies, since many GRBs are hosted by galaxies with young stellar populations \citep[e.g.,][]{Savaglio2009}.
In addition, the top-heavy initial mass function (IMF) also can increase the ratio of $\alpha$ element to iron, because theoretical predictions of elemental yields of SNe II show that [$\alpha$/Fe] increases with increasing progenitor mass \citep[e.g.,][]{Wyse1992,Woosley1995}.
The single massive star scenario might favor the top-heavy IMF of GRB host galaxies. 
Therefore, GRB host galaxies could have a high value of [O/Fe]. 
If so, the oxygen abundances of GRB hosts might not always be a good indicator of iron abundance. 


To test these scenarios, it is important to directly measure the iron abundance of the GRB host with high oxygen abundance. 
Some nearby GRB host galaxies enable us to measure the flux of the weak [Fe~{\sc iii}]$\lambda$4658 line that is necessary to estimate the iron abundance based on emission-line diagnostics.
Therefore, we measured the iron abundance of the nearby host galaxy of GRB 080517 at $z$=0.089.
The oxygen abundance of GRB 080517 is around the solar value \citep{Stanway2015}. 
This is the first attempt to measure the iron abundance of the GRB host with high oxygen abundance.


In addition, we present the iron abundance measurements of the GRB 980425 host galaxy at $z$=0.0085.
GRB 980425 has been reported to show low oxygen abundance \citep{Hammer2006} but is an ideal case to constrain the iron abundance at the position of the GRB, thanks to the close distance and high signal-to-noise ratio (S/N) of the spectrum. 
Often, iron abundance was measured using the GRB afterglow \citep[e.g.,][]{Prochaska2007}, resulting in the integrated abundance rather than at the GRB site. 
In this work, we measure iron abundance from the spectra of GRB hosts. 
As such, for GRB 980425, we are able to constrain the iron abundance of the GRB site. 
This is the first case in which the iron abundance was constrained at the GRB site without contamination from the other part of the host galaxy.
The Wolf--Rayet (WR) region in GRB 980425 host, which is $\sim$ 800 kpc away from the GRB site, is also one candidate for the birthplace of the GRB \citep{Hammer2006}.
Therefore, we include the WR spectrum in our analysis, as well as the GRB 980425 site.

Throughout this paper, we use the solar abundances by \citet{Asplund2009}.

\section{\bf Observations} 
\label{observations}
\begin{figure}
	\includegraphics[width=8.5cm]{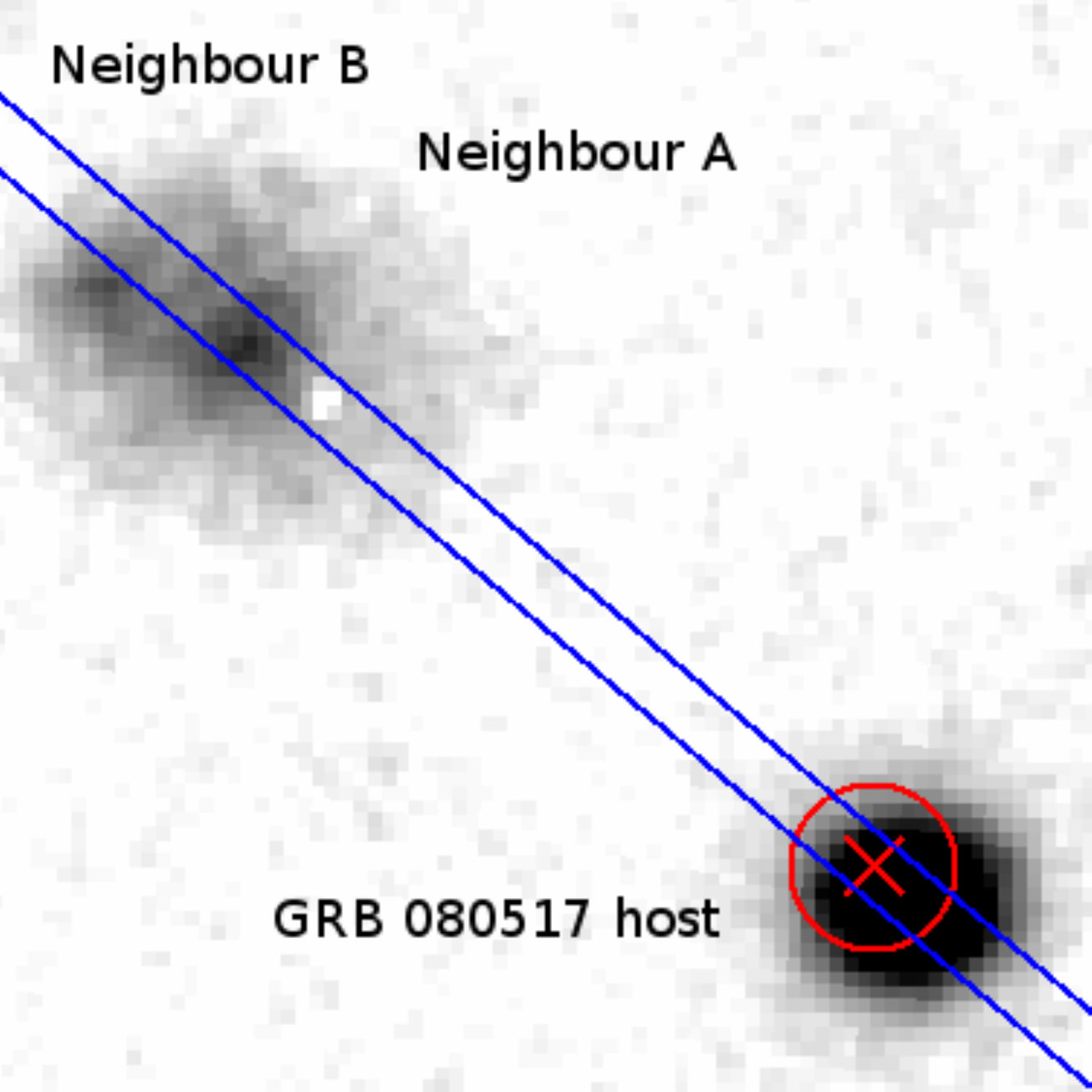}
    \caption{
    Optical image (20$\arcsec$ $\times$ 20$\arcsec$) of the host galaxy of GRB 080517 and companion galaxies obtained by WHT \citep{Stanway2015}.
    North is up, and east is left.
    The slit position of Subaru/FOCAS is indicated by blue lines.
    The 90\% confidence error circle (1$\arcsec$.5 in radius) from the {\it Swift} XRT detection of the afterglow is indicated by the red circle.
    The red cross is the center of the afterglow. 
    }
    \label{fig1}
\end{figure}

\begin{figure*}
    \begin{center}
	\includegraphics[width=14cm]{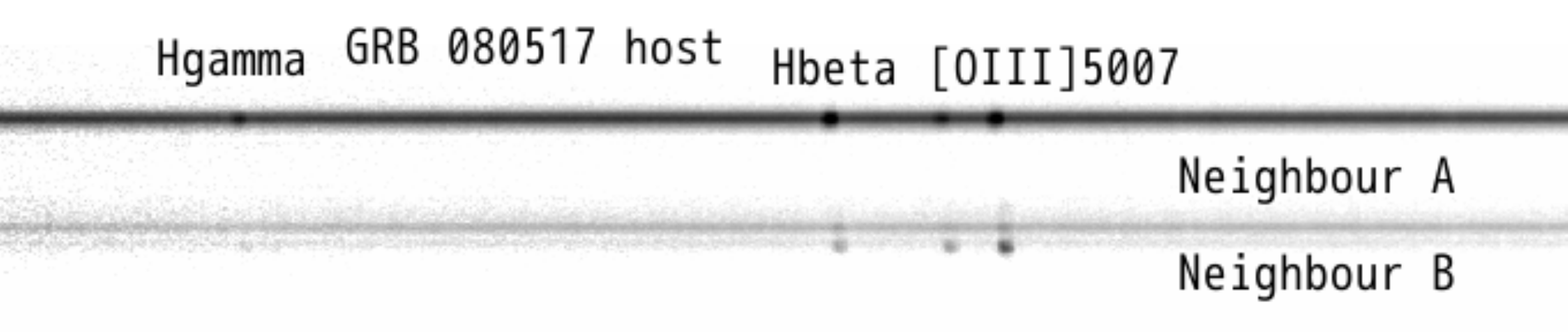}
    \caption{
    The 2D spectrum of the GRB 080517 host obtained by Subaru/FOCAS, including those of neighbor galaxies.
    }
    \label{fig2}
    \end{center}
\end{figure*}

To measure the iron abundance, we obtained a spectrum of the host galaxy of GRB 080517 using the Subaru/Faint Object Camera and Spectrograph \citep[FOCAS;][]{Kashikawa2002} with a B300 grism and L600 filter covering 4000--6000 \AA\, as well as the standard star SAO 025976 on 2017 January 8.
The slit was oriented to cover the central part of the host galaxy, GRB position, and neighbor galaxy, as shown in Figure \ref{fig1}. 
The slit width of 1$\arcsec$.0 used here corresponds to a resolving power of $R \sim$ 400.
Six iterations of three dithering exposures were performed to achieve in a total of 3 hr of exposure on source.
We reduced 2-dimensional (2D) spectral data in a standard procedure using IRAF.
Figure \ref{fig2} shows clear detections of strong emission lines such as H$\beta$ and [O~{\sc iii}]$\lambda$5007 and continuum of the host galaxy.
The central 5 pixels, which correspond to 0$\arcsec$.52 on the sky and 0.9 kpc in physical distance on the host, are extracted to make a 1D spectrum of the host.

In addition, we investigate the 1D spectra of the GRB 980425 host galaxy extracted from the burst position (SN) and WR region that is $\sim$800 kpc away from the burst position, which were reduced by \citet{Hammer2006}.
The spectra were obtained with the VLT/FORS2 (600B and 600RI grisms with a resolution $R\sim$ 1300) covering 3322-8624 \AA\ \citep{Hammer2006}.

The detection or constraint of the emission-line flux of iron is essential to estimate iron abundance (see details in section \ref{abundance_determination}).
In the previous study, the GRB 980425 host was also observed with VLT/MUSE ranging from 4750 to 9300 \AA, which covers [Fe~{\sc iii}]$\lambda$4881 \citep{Kruhler2017}.
We note, however, that [Fe~{\sc iii}]$\lambda$4881 covered by MUSE is empirically weaker than [Fe~{\sc iii}]$\lambda$4658 covered by FORS2 by a factor of $\sim$3 \citep{Rodriguez2002}.
The exposure time in VLT/MUSE is 2 times longer than that in VLT/FORS2.
In a simple estimate, we expect that the S/N in MUSE is $\sqrt{2}$ times higher than the FORS2 spectrum for the same emission-line flux.
The [Fe~{\sc iii}]$\lambda$4658 in the FORS2 spectrum should have a higher S/N ratio than the [Fe~{\sc iii}]$\lambda$4881 in MUSE by factor of $\sim$3/$\sqrt{2}$.
No clear feature of [Fe~{\sc iii}]$\lambda$4881 can be found in the MUSE spectrum reported in Figure 2 in \citet{Kruhler2017}.
Therefore, we use VLT/FORS2 data rather than VLT/MUSE in our analysis.

\begin{figure}
	\includegraphics[width=8.5cm]{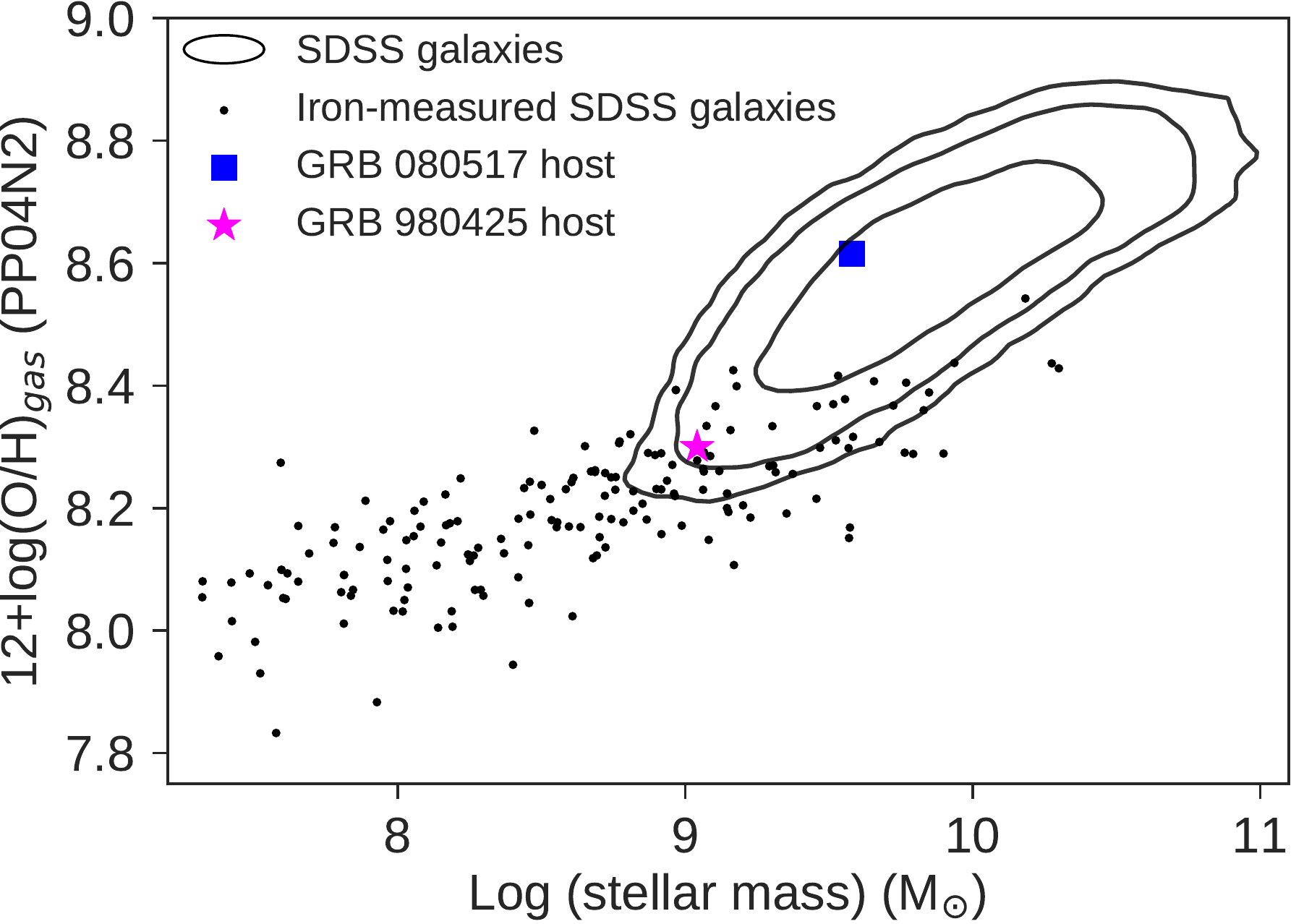}
    \caption{
   Stellar-mass metallicity relation of star-forming galaxies.
   The three black contours contain 68.3\%, 95.5\%, and 99.7\% of star-forming galaxies selected from the SDSS DR7.
   The SDSS galaxies with a measurement of iron abundance \citep{Izotov2006} are indicated by black dots.
   Colored symbols are the GRB 080517 and 980425 host galaxies.
   The oxygen abundance of the GRB host is measured from emission-line fluxes integrated over the whole host galaxies \citep{Christensen2008,Stanway2015} based on the metallicity calibration by \cite{Pettini2004}.
    }
    \label{fig3}
\end{figure}

The reference galaxies to the GRB hosts are collected from the Sloan Digital Sky Survey (SDSS), in which gas-phase iron abundances are measured based on the Te method \citep{Izotov2006}.
These galaxies are likely biased toward low metallicity, because not only weak [Fe~{\sc iii}]$\lambda$4658 but also [O~{\sc iii}]$\lambda$4363 need to be detected.
The star-forming galaxies with strong emission lines are biased toward higher star-forming rate and lower metallicity \citep[e.g.,][]{Mannucci2010} with young stellar populations \citep{Izotov2006}.
Figure \ref{fig3} shows the stellar mass-metallicity relation of normal star-forming galaxies (black contours) selected from the SDSS Data Release 7 \citep[DR7;][]{Abazajian2009} at a redshift between 0.04 and 0.1 \citep[see also][for details of sample selections]{Hashimoto2018}.
We used the publicly available catalogue of emission-line fluxes and stellar mass produced in collaboration between the Max Planck Institute for Astrophysics and Johns Hopkins University \citep{Kauffmann2003,Brinchmann2004,Salim2007}.
We note that the oxygen abundance in Figure \ref{fig3} is based on the strong emission diagnostic \citep{Pettini2004}, because the Te method is not available for the majority of SDSS galaxies.
The oxygen abundances of iron-measured SDSS galaxies (black dots) are actually comparable to the low-metallicity end of the stellar mass-metallicity relation of normal star-forming galaxies.
The oxygen abundances of the GRB 080517 and 980425 hosts are calculated from the strong emission-line fluxes integrated over the whole host galaxies reported in \citet{Christensen2008} and \citet{Stanway2015}.
We used stellar masses of GRB hosts calculated in the literature \citep{Castro2010,Stanway2015}.
The oxygen abundance of the GRB 080517 host is well above the distribution of iron-measured SDSS galaxies and consistent with the stellar mass-metallicity relation of normal star-forming galaxies.
The GRB 980425 host is between normal star-forming galaxies and iron-measured SDSS galaxies.

\section{Gas-phase abundance determination} 
\label{abundance_determination}
We follow the formulations of the gas-phase oxygen, neon, sulfur, and iron abundances by \citet{Izotov2006}, which can be calculated from observed emission lines.
Here the neon and sulfur abundances are required to estimate the depletion factor of iron discussed in section \ref{total_abundance}.
In \citet{Izotov2006}, the oxygen abundance is based on the so-called Te method, which is the direct accumulation of O$^{+}$/H$^{+}$ and O$^{2+}$/H$^{+}$ containing $\sim$ 99\% of gas-phase oxygen, i.e.,
\begin{equation}
\frac{\rm O}{\rm H}=\frac{{\rm O}^{+}}{{\rm H}^{+}}+\frac{{\rm O}^{2+}}{{\rm H}^{+}}.
\end{equation}
According to \citet{Izotov2006}, O$^{+}$/H$^{+}$ and O$^{2+}$/H$^{+}$ can be expressed as 
\begin{equation}
\label{eq2}
\begin{split}
12+\log {\rm O}^{+}/{\rm H}^{+}&=\log \frac{[{\rm OII}]\lambda 3727}{{\rm H}\beta}+5.961+\frac{1.676}{t}\\
&\quad -0.40\log t -0.034t+\log (1+1.35x),
\end{split}
\end{equation}
\begin{equation}
\label{eq3}
\begin{split}
12+\log {\rm O}^{2+}/{\rm H}^{+}&=\log \frac{[{\rm OIII}]\lambda 4959+[{\rm OIII}]\lambda 5007}{{\rm H}\beta}+6.200\\
&\quad +\frac{1.251}{t}-0.55\log t -0.014t,
\end{split}
\end{equation}
where $t=10^{-4}T_{e}$(O~{\sc iii}).
The electron temperature, $t$, can be derived from the equation
\begin{equation}
t=\frac{1.432}{\log [([{\rm OIII}]\lambda 4959+[{\rm OIII}]\lambda 5007)/[{\rm OIII}]\lambda 4363]-\log C_{T}},
\end{equation}
where 
\begin{equation}
C_{T}=(8.44-1.09t+0.5t^{2}-0.08t^{3})\frac{1+0.0004x}{1+0.044x},
\end{equation}
and $x=10^{-4}N_{e}t^{-0.5}$.
The electron density, $N_{e}$, is derived from the [S~{\sc ii}]$\lambda$6717/[S~{\sc ii}]$\lambda$6731 line ratio \citep[e.g.,][]{Osterbrock1989,Proxauf2014}.
The abundances of neon, sulfur, and iron are calculated from the abundances of the specific ionized elements and ionization correction factors (ICFs),
\begin{equation}
12+\log {\rm Ne/H}=12+\log {\rm Ne}^{2+}/H^{+}+\log ICF({\rm Ne}^{2+}),
\end{equation}
where 
\begin{equation}
\label{eq7}
\begin{split}
12+\log {\rm Ne}^{2+}/{\rm H}^{+}&=\log \frac{[{\rm NeIII}]\lambda 3869}{{\rm H}\beta}+6.444\\
&\quad +\frac{1.606}{t}-0.42\log t - 0.009t,
\end{split}
\end{equation}
\begin{equation}
\begin{split}
ICF({\rm Ne}^{2+})&=-0.385w+1.365+0.022/w, {\rm low~Z},\\
&=-0.405w+1.382+0.021/w, {\rm intermed.~Z},\\
&=-0.591w+0.927+0.546/w, {\rm high~Z},
\end{split}
\end{equation}
\begin{equation}
w={\rm O}^{2+}/({\rm O}^{+}+{\rm O}^{2+}),
\end{equation}
and
\begin{equation}
12+\log {\rm S/H}=12+\log [{\rm S}^{+}/H^{+}+{\rm S}^{2+}/{\rm H}^{+}]+\log ICF({\rm S}^{+}+{\rm S}^{2+}),
\end{equation}
where
\begin{equation}
\label{eq11}
\begin{split}
12+\log {\rm S}^{+}/{\rm H}^{+}&=\log \frac{[{\rm SII}]\lambda 6717+[{\rm SII}]\lambda 6731}{{\rm H}\beta}+5.439\\
&\quad +\frac{0.929}{t}-0.28\log t\\
&\quad -0.018t+\log (1+1.39x),
\end{split}
\end{equation}
\begin{equation}
\label{eq12}
\begin{split}
12+\log {\rm S}^{2+}/{\rm H}^{+}&=\log \frac{[{\rm SIII}]\lambda 6312}{{\rm H}\beta}+6.690\\
&\quad +\frac{1.678}{t}-0.47\log t-0.010t,
\end{split}
\end{equation}
\begin{equation}
\begin{split}
ICF({\rm S}^{+}+{\rm S}^{2+})&=0.121v+0.511+0.161/v, {\rm low~Z},\\
&=0.155v+0.849+0.062/v, {\rm intermed.~Z},\\
&=0.178v+0.610+0.153/v, {\rm high~Z},
\end{split}
\end{equation}
\begin{equation}
v={\rm O}^{+}/({\rm O}^{+}+{\rm O}^{2+}),
\end{equation}
and
\begin{equation}
12+\log {\rm Fe/H}=12+\log {\rm Fe}^{2+}/{\rm H}^{+}+\log ICF({\rm Fe}^{2+}),
\end{equation}
where
\begin{equation}
\label{eq16}
\begin{split}
12+\log {\rm Fe}^{2+}/{\rm H}^{+}&=\log \frac{[{\rm FeIII}]\lambda 4658}{{\rm H}\beta}+6.498\\
&\quad +\frac{1.298}{t}-0.48\log t,
\end{split}
\end{equation}
and
\begin{equation}
\begin{split}
ICF({\rm Fe}^{2+})&=0.036v-0.146+1.386/v, {\rm low~Z},\\
&=0.301v-0.259+1.367/v, {\rm intermed.~Z},\\
&=-1.377v+1.606+1.045/v, {\rm high~Z}.
\end{split}
\end{equation}
The ICF here is a correction term applied to an abundance of a specific transition of ion in order to take into account the ion abundances in the other transitions \citep{Izotov2006}, i.e.,\\ 12+log($M$/H)=12+log($M_{ion}$/H)+log(ICF($M_{ion}$)).
The formulae of ICFs, categorized into \lq\lq low Z,\rq\rq \lq\lq intermed. Z,\rq\rq and \lq\lq high Z,\rq\rq are applicable to $12+\log {\rm O/H} \le 7.4$, $7.4 < 12+\log {\rm O/H}< 7.9$, and $12+\log {\rm O/H} \ge 7.9$, respectively.
In calculations of gas-phase Ne/H, S/H, and Fe/H, only one or two transitions of each element are available in the rest-frame optical spectrum.
Abundances of other transitions such as Ne and Ne$^{+}$ are counted by including ICFs.
We note that $T_{e}$(O~{\sc ii}) is adopted for the calculations of O$^{+}$, S$^{+}$, and Fe$^{2+}$ abundances in Equations (\ref{eq2}), (\ref{eq11}), and (\ref{eq16}).
Here $T_{e}$(S~{\sc iii}) is adopted for the S$^{2+}$ calculation in Equation (\ref{eq12}), and $T_{e}$(O~{\sc iii}) is used for the O$^{2+}$ and Ne$^{2+}$ calculations in Equations (\ref{eq3}) and (\ref{eq7}).
The $T_{e}$(O~{\sc ii}) and $T_{e}$(S~{\sc iii}) are derived from $T_{e}$(O~{\sc iii}) as follows \citep{Izotov2006}:
\begin{equation}
\label{eq18}
\begin{split}
t({\rm O~II})&=-0.577+t\times(2.065-0.498t), {\rm low~Z},\\
&=-0.744+t\times(2.338-0.610t), {\rm intermed.~Z},\\
&=2.967+t\times(-4.797+2.827t), {\rm high~Z},
\end{split}
\end{equation}
\begin{equation}
\begin{split}
t({\rm S~III})&=-1.085+t\times(2.320-0.420t), {\rm low~Z},\\
&=-1.276+t\times(2.645-0.546t), {\rm intermed.~Z},\\
&=1.653+t\times(-2.261+1.605t), {\rm high~Z},
\end{split}
\end{equation}
where $t({\rm O~II})=10^{-4}T_{e}$(O~{\sc ii}), $t({\rm S~III})=10^{-4}T_{e}$(S~{\sc iii}), and $t=10^{-4}T_{e}$(O~{\sc iii}).

In summary, gas-phase elemental abundances are calculated from specific emission lines appearing in the rest-frame optical wavelength (e.g., [Ne~{\sc iii}]$\lambda$3869 and [Fe~{\sc iii}]$\lambda$4658) and physical conditions of ionized gas, i.e., electron temperature, density, $\nu={\rm O}^{+}/({\rm O}^{+}+{\rm O}^{2+})$, and $w={\rm O}^{2+}/({\rm O}^{+}+{\rm O}^{2+})$.

\section{Emission-line measurements} 
\subsection{Subtraction of the stellar component}
\label{Subtraction of the stellar component}
\begin{figure*}
    \begin{center}
	\includegraphics[width=14cm]{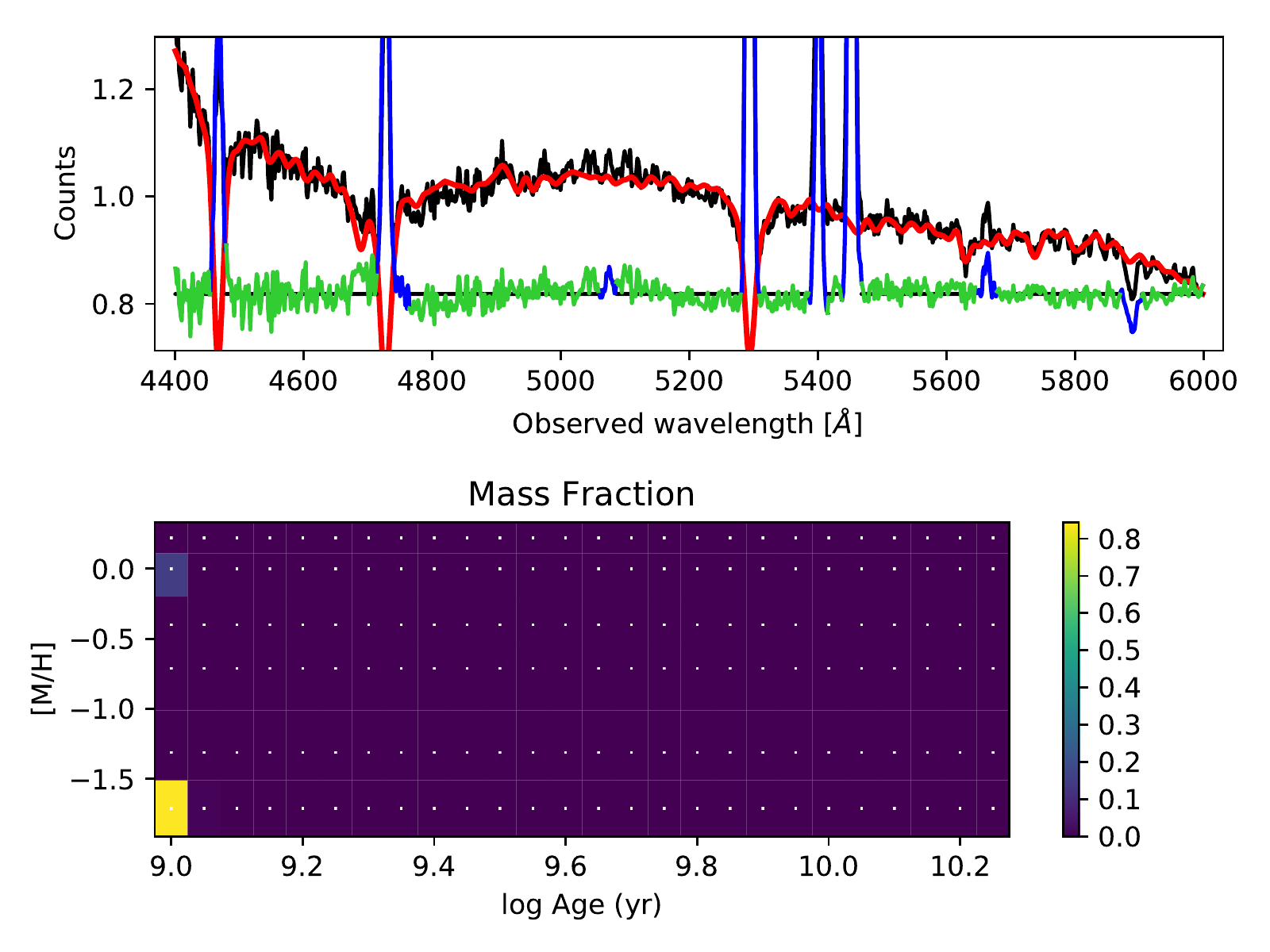}
    \caption{
    Observed 1D spectrum of the GRB 080517 host galaxy obtained by Subaru/FOCAS (top) and the best-fit fractional distribution of single stellar populations in the age-metallicity parameter space (bottom).
    Black and red lines are the observed and best-fit model spectra, respectively.
    Masked regions are marked by blue lines. 
    The residual of the subtraction of the best-fit stellar component from the observed spectrum is indicated by the green line.
    The color bar in the bottom panel displays relative mass fractions of individual single stellar populations with different ages and metallicities that compose the best-fit model spectrum of the stellar component (red line in the top panel).
    }
    \label{fig4}
    \end{center}
\end{figure*}

\begin{figure*}
    \begin{center}
	\includegraphics[width=14cm]{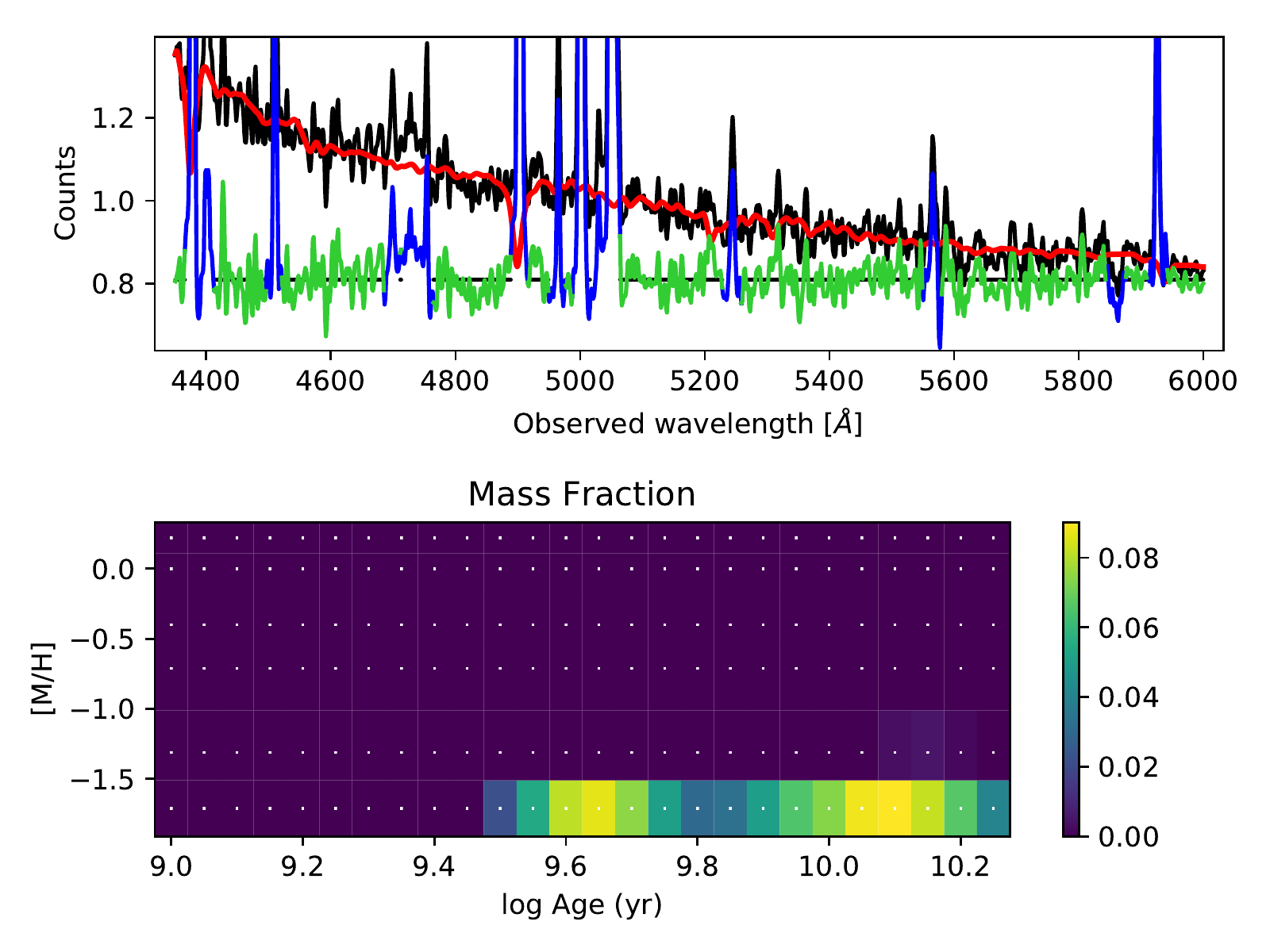}
    \caption{
    Same as Figure \ref{fig4} except for the the spectrum at the WR region of the GRB 980425 host \citep{Hammer2006}.
    }
    \label{fig5}
    \end{center}
\end{figure*}

\begin{figure*}
    \begin{center}
	\includegraphics[width=14cm]{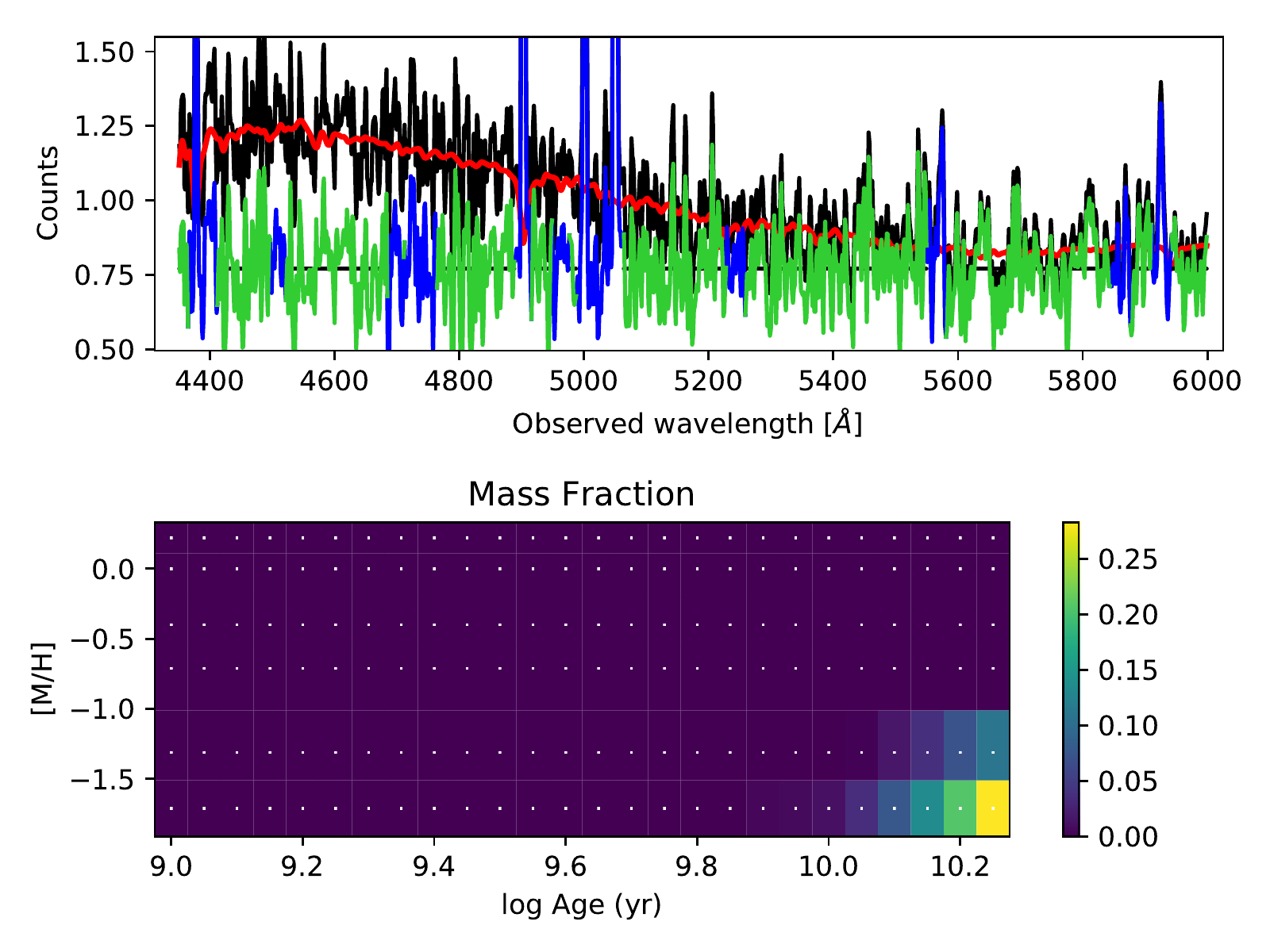}
    \caption{
    Same as Figure \ref{fig4} except for the the spectrum at the position of GRB 980425 \citep{Hammer2006}.
    }
    \label{fig6}
    \end{center}
\end{figure*}

As mentioned in section \ref{abundance_determination}, the measurements of weak emission-line fluxes (e.g., [O~{\sc iii}]$\lambda$4363 and [Fe~{\sc iii}]$\lambda$4658) are required to calculate the gas-phase elemental abundances.
These weak emission lines can be affected by the underlying stellar features.
Therefore, we performed the spectral-fitting analysis for the stellar component based on the software of penalized pixel fitting \citep[pPXF;][]{Cappellari2004,Cappellari2017}, along with the MILES stellar library \citep{Vazdekis2010}. 
The pPXF allows one to construct the best-fit stellar component consisting of single stellar populations with different ages, metallicities, and weights.
In the fitting process, the wavelength ranges of the prominent emission lines and atmospheric absorption line are masked to avoid contamination from emission/nonstellar absorption features.
The top panel of Figure \ref{fig4} shows the observed 1D spectrum of the GRB 080517 host (black line) extracted from the central 5 pixels and best-fit stellar component (red line).
We subtracted the best-fit stellar component from the observed spectrum.
The residual spectrum is indicated by the green line.
The best-fit stellar component includes single stellar populations with different ages, metallicities, and weights, as shown in the bottom panel of Figure \ref{fig4}.
Figure \ref{fig5} and \ref{fig6} are the same as Figure \ref{fig4} except for the spectrum at the WR region of the GRB 980425 host and the explosion site \citep{Hammer2006}.

On the uncertainty of the subtraction, \citet{Onodera2012} performed a similar analysis by using pPXF to construct the stellar components in low-resolution spectra ($R\sim$500) of passive galaxies.
They also successfully reproduced the observed spectra and reported that the best-fit model is indistinguishable from that from another method with a different fitting code and different stellar library, namely STARLIGHT \citep{Fernandes2005,Fernandes2009}.
The uncertainties of the stellar-continuum fitting in the GRB hosts could be within the dispersion of the residual spectra in Figure \ref{fig7}-\ref{fig9}.
Therefore, we use the residual spectra that potentially contain remaining stellar features to estimate the 3 $\sigma$ upper limit and errors of the weak emission-line fluxes.
In this sense, the uncertainty of the stellar fitting is already, to some extent, reflected in the line detection and flux errors.
Note that the strong emission lines in Figure \ref{fig4}-\ref{fig6} are not saturated.

\subsection{Emission-line fitting} 
\label{emission_fitting}
\begin{figure*}
    \begin{center}
	\includegraphics[width=12cm]{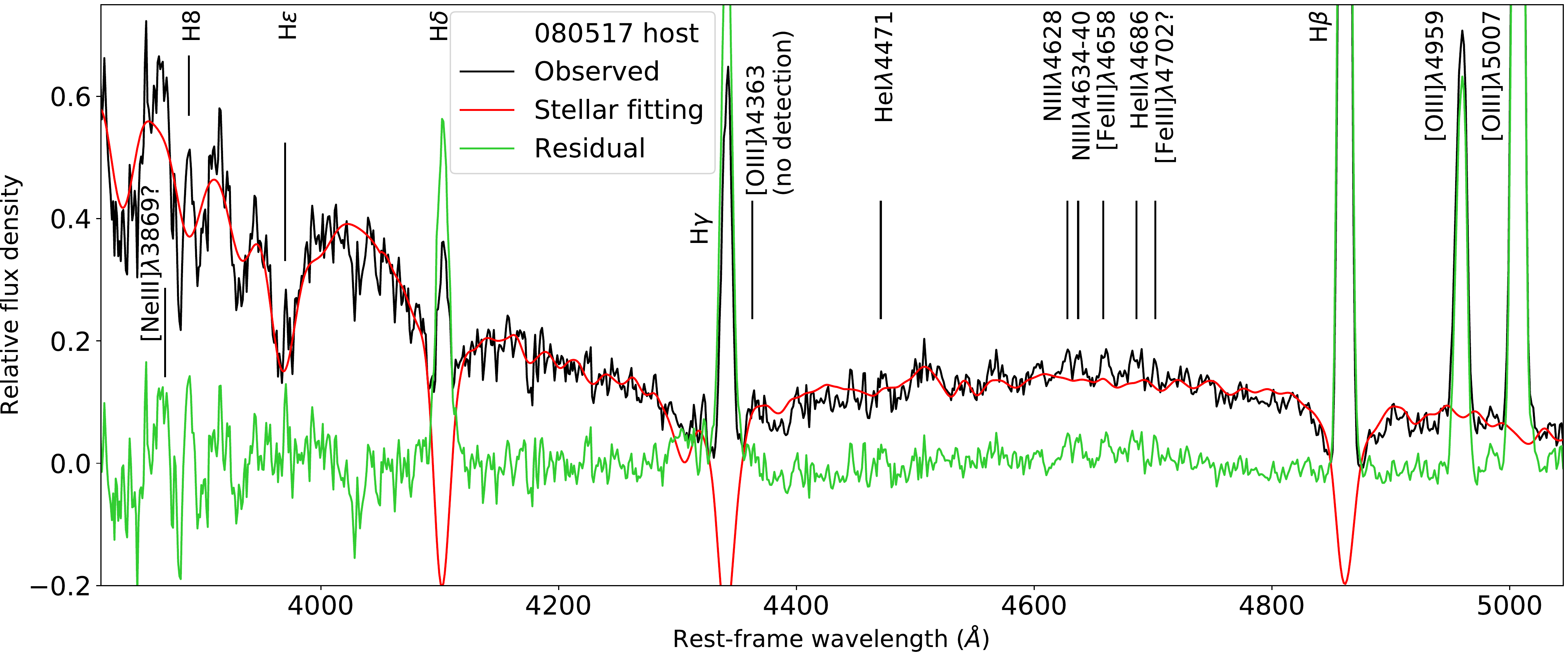}
    \caption{
    Emission-line identification in the residual spectrum of the GRB 080517 host galaxy.
    The the best-fit stellar component (red) is subtracted from the observed spectrum (black), resulting in the residual spectrum (green).
    A constant value is added to the observed spectrum and best-fit stellar component.
    }
    \label{fig7}
    \end{center}
\end{figure*}

\begin{figure*}
    \begin{center}
	\includegraphics[width=12cm]{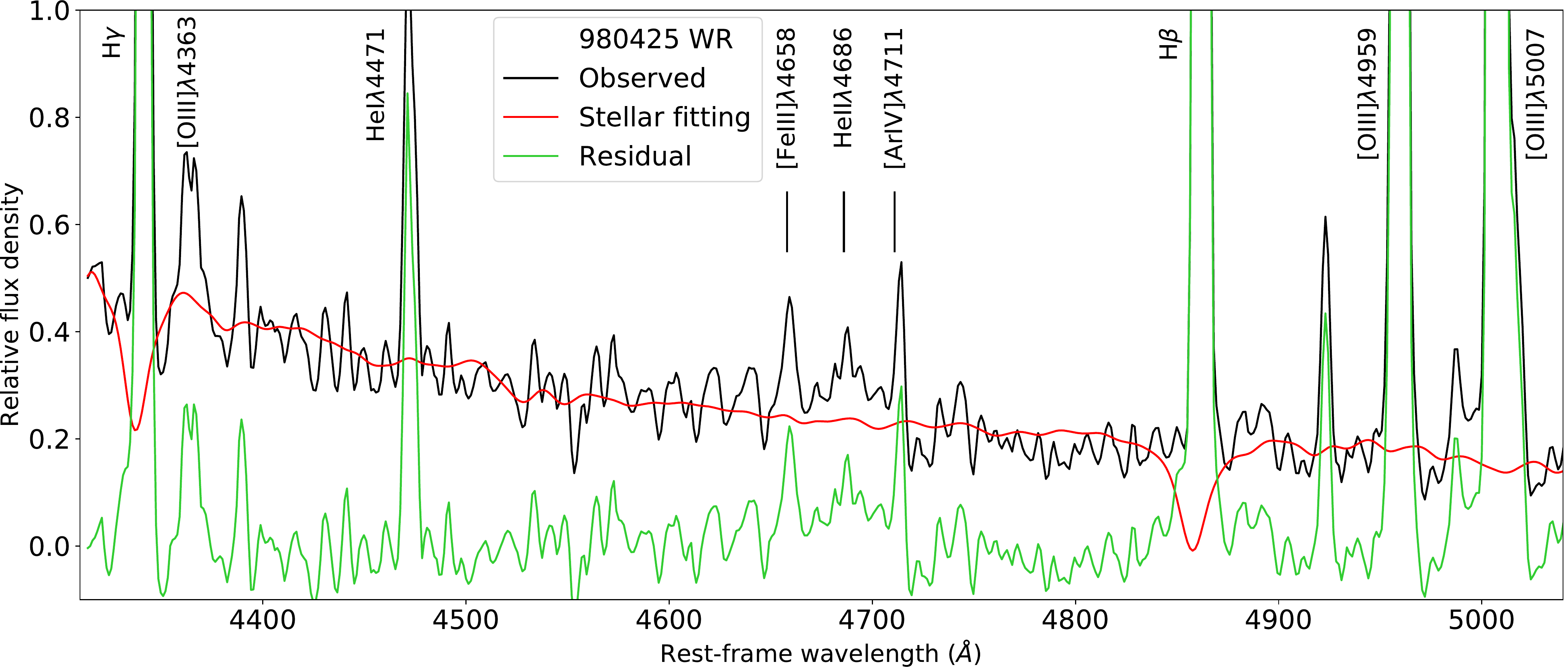}
    \caption{
    Same as Figure \ref{fig7} except for the spectrum at the WR region of the GRB 980425 host.
    }
    \label{fig8}
    \end{center}
\end{figure*}

\begin{figure*}
    \begin{center}
	\includegraphics[width=12cm]{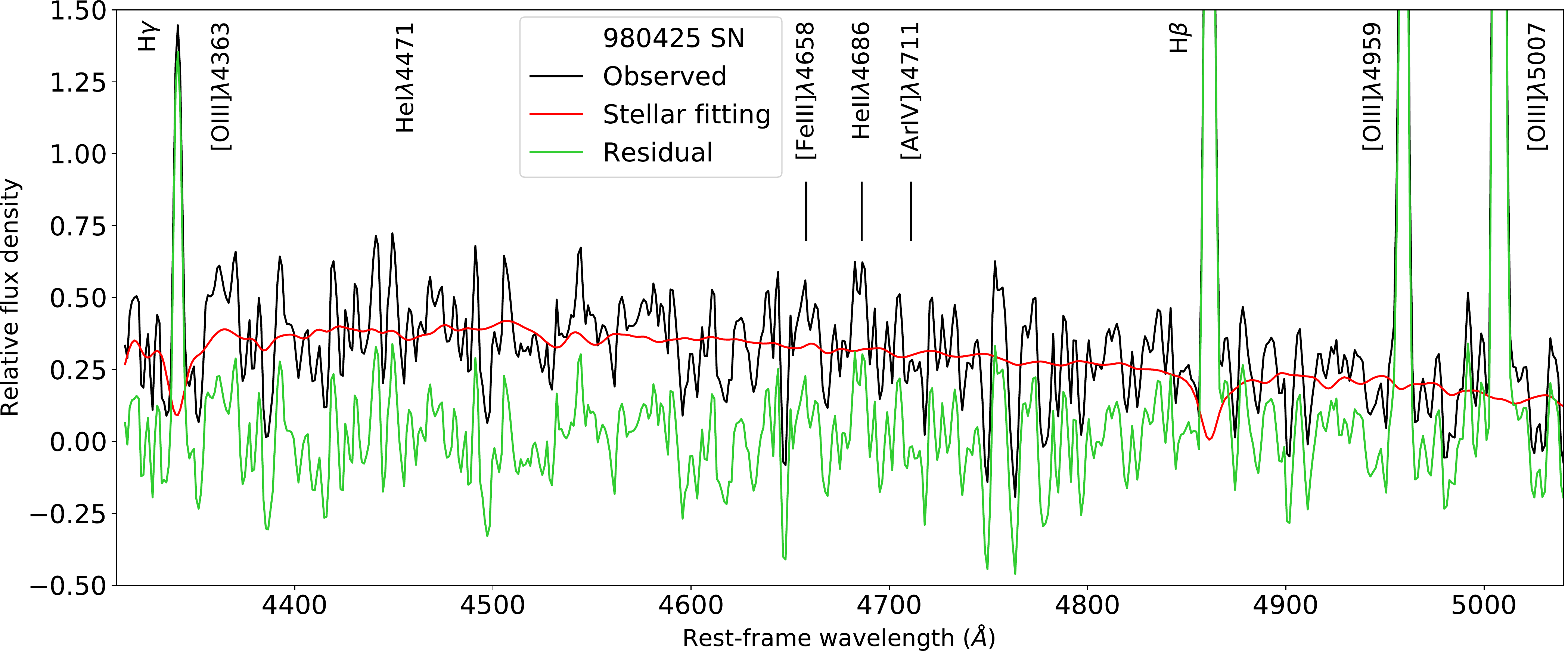}
    \caption{
    Same as Figure \ref{fig7} except for the spectrum at the site of GRB 980425.
    }
    \label{fig9}
    \end{center}
\end{figure*}

The residual spectra of the GRB 080517 and 980425 hosts are expanded in Figure \ref{fig7}-\ref{fig9}.
We successfully detected [Fe~{\sc iii}]$\lambda$4658 in the spectra of the GRB 080517 host and the WR region of the GRB 980425 host.
The possible effect of an underlying stellar component on the [Fe~{\sc iii}]$\lambda$4658 line is likely small, because there is no strong stellar absorption feature on the wavelength of the [Fe~{\sc iii}]$\lambda$4658 line in the best-fit stellar spectra shown as red lines in Figure \ref{fig7} and \ref{fig8}.
We have checked possible detections of other Fe lines in the spectra of GRB 080517 and 980425 and found no clear features.
The second-strongest Fe line is [Fe~{\sc iii}]$\lambda$4881, which is typically weaker than [Fe~{\sc iii}]$\lambda$4658 by factor of $\sim$3.
The expected S/N of [Fe~{\sc iii}]$\lambda$4881 can be estimated from that of [Fe~{\sc iii}]$\lambda$4658 and is lower than 3 in all GRB host spectra.
This is consistent with the nondetection of other Fe lines.

In the spectrum of the GRB 080517 host (Figure \ref{fig7}), the observed feature around [O~{\sc iii}]$\lambda$4363 is consistent with the stellar component without any emission-line features.
The possible emission-line feature around [Ne~{\sc iii}]$\lambda$3869 is strongly affected by the stellar component.
Therefore, we use 3 $\sigma$ upper limits of these undetected emission lines in further analysis in section \ref{gas-phase_abundance} and \ref{total_abundance}.

The spectrum at the WR region of the GRB 980425 host has quite a high S/N.
Actually, [O~{\sc iii}]$\lambda$4363 is clearly detected in our residual spectrum of GRB 980425, as shown in Figure \ref{fig8}.
The 3$\sigma$ upper limit of the weak emission line is estimated to be 0.05 erg s$^{-1}$ cm$^{-2}$ \AA$^{-1}$ in the residual spectrum by assuming the FWHM of 5.6 \AA\ measured for H$\beta$.
This value is almost half of the 3$\sigma$ (0.09 erg s$^{-1}$ cm$^{-2}$ \AA$^{-1}$) reported in \citet{Hammer2006}.
This is probably because they did not subtract the stellar component in measuring the emission-line fluxes. 
The stellar component can affect not only on the flux measurements of weak emission lines but also the error estimates. 
The noise estimate around the emission line in the 1D spectrum before the stellar subtraction is contaminated by the stellar features (e.g., zigzag pattern but not true noise), which causes an overestimate of the data dispersion and noise estimate.

The stellar features likely cause the overestimates of the errors around the weak emission lines.
We assumed a conservative value of S/N=50 for strong emission lines such as [O~{\sc ii}]$\lambda$3727, H$\alpha$ and [O~{\sc iii}]$\lambda$5007, rather than adopting a 1$\sigma$ error of $\sim$0.016 erg s$^{-1}$ cm$^{-2}$ \AA$^{-1}$.

The S/N of the spectrum at the site of GRB 980425 is relatively poorer than that of the WR region (Figure \ref{fig9}).
No [Fe~{\sc iii}]$\lambda$4658 and [O~{\sc iii}]$\lambda$4363 are clearly detected.
We use a 3$\sigma$ upper limit on [Fe~{\sc iii}]$\lambda$4658 to constrain the iron abundance at the explosion site of GRB 980425.
Because the uncertainty of $T_{e}$(O~{\sc iii}) implied from [O~{\sc iii}]$\lambda$4363 is large, we use $T_{e}$(O~{\sc ii}) and equation \ref{eq18} to estimate $T_{e}$(O~{\sc iii}) at the GRB site.
Here $T_{e}$(O~{\sc ii}) is calculated from [O~{\sc ii}]$\lambda$3726+3729/[O~{\sc ii}]$\lambda$7320+7330 \citep{Hammer2006} by using the nebular task in IRAF.
We assumed a conservative value of S/N=10 for strong emission lines.

\begin{figure}
	\includegraphics[width=8.5cm]{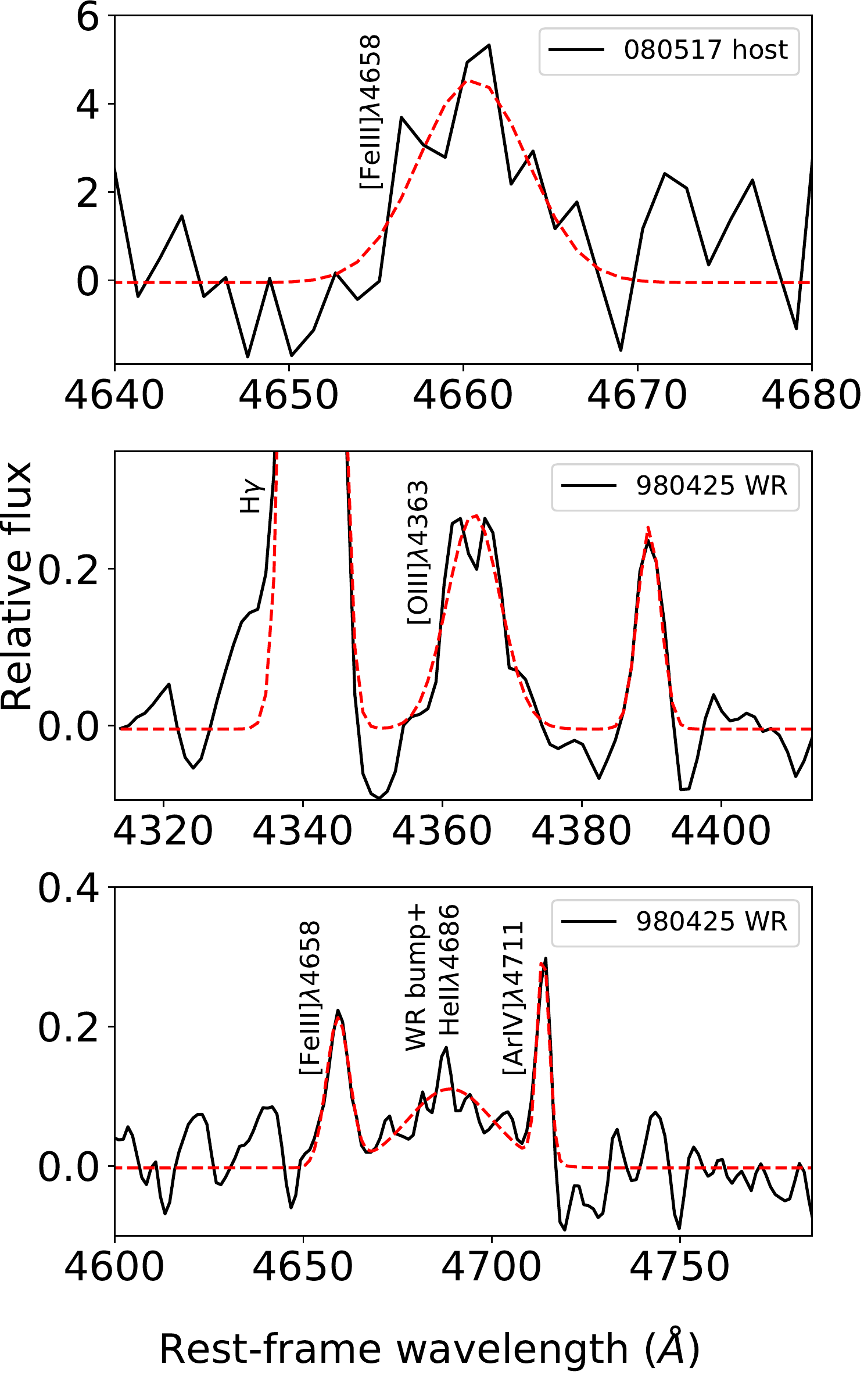}
    \caption{
    Emission-line fitting of [O~{\sc iii}]$\lambda$4363 and [Fe~{\sc iii}]$\lambda$4658 with the gaussian profile and constant base line, indicated by red dashed lines. 
    Black lines are the residual spectra between the observed spectra and best-fit stellar components.
    }
    \label{fig10}
\end{figure}

In combination with the strong emission-line measurements reported by \citet{Hammer2006}, we detected all of the emission lines of the WR region of the GRB 980425 host required to calculate the elemental abundances mentioned in section \ref{abundance_determination}.
The emission lines of [O~{\sc iii}]$\lambda$4363 and [Fe~{\sc iii}]$\lambda$4658 detected in the spectra are fitted with the single gaussian function and constant base line. 
In the spectrum of GRB 980425, the base lines underlying [O~{\sc iii}]$\lambda$4363 and [Fe~{\sc iii}]$\lambda$4658 lines can be affected by neighboring emission lines. 
Therefore, we simultaneously fit [O~{\sc iii}]$\lambda$4363 and [Fe~{\sc iii}]$\lambda$4658 along with nearby emission lines, as shown in Figure \ref{fig10}.
Although we set the base line as a free parameter, the value is actually almost identical to 0.0.

We summarize the measured emission-line fluxes in the spectra of the GRB 080517 and 980425 hosts in table \ref{tab1} along with the strong emission-line fluxes reported in the literature \citep{Hammer2006,Stanway2015}.

\begin{table}
    \caption{
     Emission-line Fluxes Relative to the H$\beta$ of Our Sample.
    }
    \label{tab1}
     \begin{flushleft}
     \begin{tabular}{|l|l|l|l|}\hline \hline
     ID&080517 Host&980425 WR&980425 SN\\ \hline \relax
     [O~{\sc ii}]$\lambda$3727&5.530$\pm$0.532$^{b}$&1.243$^{c}\pm$0.025$^{d}$&4.418$^{c}\pm$0.4418$^{e}$\\ \relax
     [Ne~{\sc iii}]$\lambda$3869&0.032$^{a}$ (3$\sigma$)&0.319$^{c}\pm$0.006$^{d}$&0.604$^{c}\pm$0.060$^{e}$\\ \relax
     [O~{\sc iii}]$\lambda$4363&0.007$^{a}$ (3$\sigma$)&0.043$\pm$0.006$^{a}$&0.055$^{a}$ (3$\sigma$)\\ \relax
     [Fe~{\sc iii}]$\lambda$4658&0.012$\pm$0.003$^{a}$&0.027$\pm$0.004$^{a}$&0.055$^{a}$ (3$\sigma$)\\ \relax
     H$\beta^{g}$&1.000$\pm$0.060$^{b}$&1.000$^{c}\pm$0.020$^{d}$&1.000$^{c}\pm$0.100$^{e}$\\ \relax
     [O~{\sc iii}]$\lambda$4959&0.345$\pm$0.018$^{b}$&1.404$^{c}\pm$0.028$^{d}$&0.989$^{c}\pm$0.099$^{e}$\\ \relax
     [O~{\sc iii}]$\lambda$5007&1.018$\pm$0.060$^{b}$&5.684$^{c}\pm$0.114$^{d}$&2.242$^{c}\pm$0.224$^{e}$\\ \relax
     [S~{\sc iii}]$\lambda$6312&0.036$^{a}$ (3$\sigma$)&0.025$^{c}\pm$0.001$^{c}$&0.044$^{c}\pm$0.015$^{f}$\\ \relax
     H$\alpha$&6.137$\pm$0.143$^{b}$&5.195$^{c}\pm$0.104$^{d}$&4.242$^{c}\pm$0.424$^{e}$\\ \relax
     [S~{\sc ii}]$\lambda$6717&0.792$\pm$0.030$^{b}$&0.308$^{c}\pm$0.006$^{d}$&0.659$^{c}\pm$0.066$^{e}$\\ \relax
     [S~{\sc ii}]$\lambda$6731&0.774$\pm$0.030$^{b}$&0.243$^{c}\pm$0.005$^{d}$&0.923$^{c}\pm$0.092$^{e}$\\ \hline
     \end{tabular}\\
     $^{a}$Our measurements.\\
     $^{b}$Fluxes and errors reported by \citet{Stanway2015} are converted to those relative to H$\beta$.\\
     $^{c}$Fluxes reported by \citet{Hammer2006} are converted to those relative to H$\beta$.\\
     $^{d}$The conservative value of S/N=50 is assumed.\\
     $^{e}$The conservative value of S/N=10 is assumed.\\
     $^{f}$S/N=3 is assumed.\\
     $^{g}$H$\beta$ fluxes are in units of $2.52\times 10^{-15}$, $3.54\times 10^{-15}$, and $9.0\times 10^{-17}$erg s$^{-1}$ cm$^{-2}$ for the 080517 host, 980425 WR, and SN, respectively \citep{Hammer2006,Stanway2015}.\\
     \end{flushleft}
\end{table}

\section{Gas-phase abundance}
\label{gas-phase_abundance}
\begin{figure*}
    \begin{center}
	\includegraphics[width=17cm]{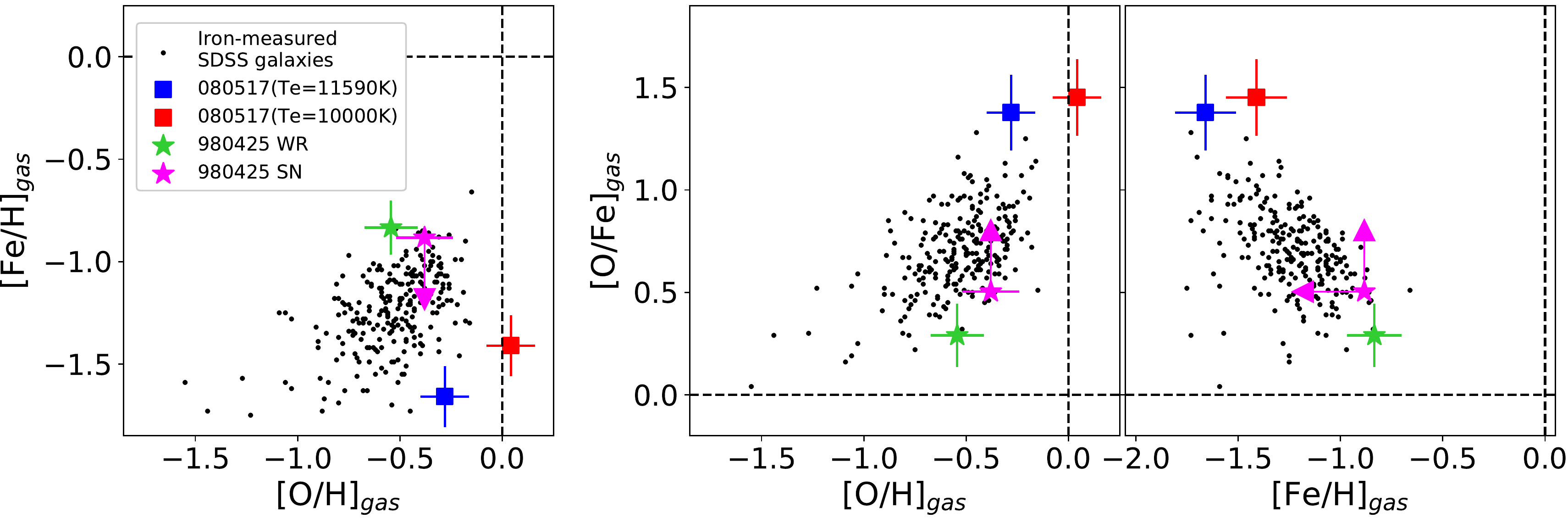}
    \caption{
    Gas-phase iron and oxygen abundances of the GRB 080517 and 980425 hosts (colored data points).
    Error bars contain observational errors of flux measurements and the statistical uncertainty of the abundance determination, i.e., $\sim$0.1 dex \citep{Stasinska2005}.
    Abundances measured in the iron-measured SDSS galaxies are indicated by black dots.
    The solar abundances are shown by dashed lines.
    }
    \label{fig11}
    \end{center}
\end{figure*}

The gas-phase iron and oxygen abundances of our sample are calculated according to the formulations in section \ref{abundance_determination}. 
We used extinction-corrected emission-line fluxes based on the extinction law by \citet{Calzetti2000} and the Balmer decrement (intrinsic H$\alpha$/H$\beta$ = 2.86).
Adopting the different extinction laws, such as the Milky Way law \citep{Seaton1979}, and different intrinsic values of H$\alpha$/H$\beta$ \citep{Osterbrock1989} has no impact on abundance measurements, because the resulting difference is smaller than the typical statistical uncertainty of the abundance measurement, $\sim 0.1$ dex \citep{Stasinska2005}.
The gas-phase metal abundances are summarized in table \ref{tab2} along with the adopted solar abundances.
Figure \ref{fig11} shows [O/H]$_{gas}$ versus [Fe/H]$_{gas}$ (left) and [O/H]$_{gas}$ (or [Fe/H]$_{gas}$) versus [O/Fe]$_{gas}$ (right).
Since the [O~{\sc iii}]$\lambda$4363 of the GRB 080517 host is not detected and $T_{e}$(O~{\sc ii}) is not available, we assumed two cases of electron temperatures of $T_{e}$(O~{\sc iii})=11,590K, corresponding to the 3$\sigma$ upper limit, and $T_{e}$(O~{\sc iii})=10,000K, which is the typical value in the H {\sc ii} region.
Note that the iron-measured SDSS galaxies are well below the solar value in both of [O/H]$_{gas}$ and [Fe/H]$_{gas}$. 
This is because the flux measurements of [O~{\sc iii}]$\lambda$4363 and [Fe~{\sc iii}]$\lambda$4658 are possible only for galaxies with strong emission lines. 
In general, the star-forming galaxies with strong emission lines are biased toward higher star-forming rates and lower metallicities \citep[e.g.,][]{Mannucci2010} with young stellar populations \citep{Izotov2006}. 
Therefore, the iron-measured SDSS galaxies collected here are not representative of the star-forming galaxies in the universe.
These galaxies are biased toward lower metallicity, as mentioned in section \ref{observations}, resulting in lower values of [O/H]$_{gas}$ and [Fe/H]$_{gas}$.

In the left panel of Figure \ref{fig11}, the gas-phase iron abundance, [Fe/H]$_{gas}$, of GRB 080517 is well below than that of the majority of the iron-measured SDSS galaxies regardless of the assumption of the electron temperature.
While the GRB 980425 host almost overlaps with the iron-measured SDSS galaxies, the upper limit on [Fe/H]$_{gas}$ at the GRB position is slightly lower than that at the WR region, suggesting the low iron abundance environment at the explosion site.

The top-heavy IMF is also expected to produce strong emission lines through intense ionizing photons from massive stars.
Thus, the iron-measured SDSS galaxies also might be biased toward the top-heavy IMF.
As mentioned in section \ref{introduction}, young stellar populations and the top-heavy IMF can increase the ratios of $\alpha$ elements to iron, such as [O/Fe].
In the right panel of Figure \ref{fig11}, as is expected from the possible bias toward young age and/or top-heavy IMF, the iron-measured SDSS galaxies show the overabundance of oxygen compared with iron, i.e., [O/Fe]$_{gas} > 0.0$, likely due to the less iron enrichment by SNe Ia and/or abundant production of oxygen relative to iron by the top-heavy IMF.
The [O/Fe]$_{gas}$ of the GRB 080517 host is larger than that of the iron-measured SDSS galaxies.
The [O/Fe]$_{gas}$ of the GRB 980425 host is comparable to that of the lowest end of the iron-measured SDSS galaxies and is still higher than the solar value.
The [O/Fe]$_{gas}$ at the GRB position is indicated to be higher than that in the WR region.

In terms of the gas-phase iron abundance, these GRB hosts are consistent with or outstanding from the iron-measured SDSS galaxies, which suggests peculiarly low iron abundances and significant overabundances of oxygen relative to iron.
However, iron is a well-known refractory element whose gas-phase abundance can be $\sim$ 1-2 orders of magnitude less than that locked onto dust grains, depending on the physical environment in the interstellar medium \citep[e.g.,][]{Oliva2001,DeCia2016}.
Therefore, the iron environment related to the mechanism of GRBs should be investigated by the total abundance, including both of gas-phase abundance and depleted abundance onto dust grains.

\section{Total abundance}
\label{total_abundance}
\subsection{Depletion factor}
The gas-phase abundance of a specific element $X$, $[X/{\rm H}]_{gas}$, can be simply expressed as 
\begin{equation}
[X/{\rm H}]_{gas}=[X/{\rm H}]_{total}+\delta_{X}
\end{equation}
where $[X/{\rm H}]_{total}$ is the total abundance of element $X$ and $\delta_{X}$ is the depletion factor, which is basically negative.
The gas-phase abundance ratio of $X$ to the $\alpha$ element, neon, is 
\begin{equation}
\label{depletion}
[X/{\rm Ne}]_{gas}=[X/{\rm Ne}]_{total}+\delta_{X},
\end{equation}
because neon is a nonrefractory element (or noble gas), i.e., $\delta_{\rm Ne} = 0.0$.
If the total abundance ratio to neon, $[X/{\rm Ne}]_{total}$, is known, $\delta_{X}$ can be estimated.
Here $[{\rm Fe/Ne}]_{total}$ is coupled with the variety of contributions from SNe Ia in the individual galaxies and possible different IMFs, especially for peculiar galaxies such as GRB hosts.
Therefore, the prediction of $[{\rm Fe/Ne}]_{total}$ is difficult in general. 

The abundance ratio of two $\alpha$ elements such as [O/Ne] and [S/Ne] is unlikely to be affected by varieties of chemical enrichment by SNe Ia and the possible different shape of IMFs, because $\alpha$ elements are generally believed to be produced by the same physical process through SN IIe.
Therefore, the abundance ratio of two $\alpha$ elements is expected to be the solar value as a first-order approximation, and the deviation from the solar value is likely dominated by the depletion factor of the $\alpha$ element \citep[e.g.,][]{Steidel2016}.
If such is the case, Equation (\ref{depletion}) is equivalent to
\begin{equation}
[\alpha/{\rm Ne}]_{gas}=\delta_{\alpha}.
\end{equation}

\begin{figure}
	\includegraphics[width=8.5cm]{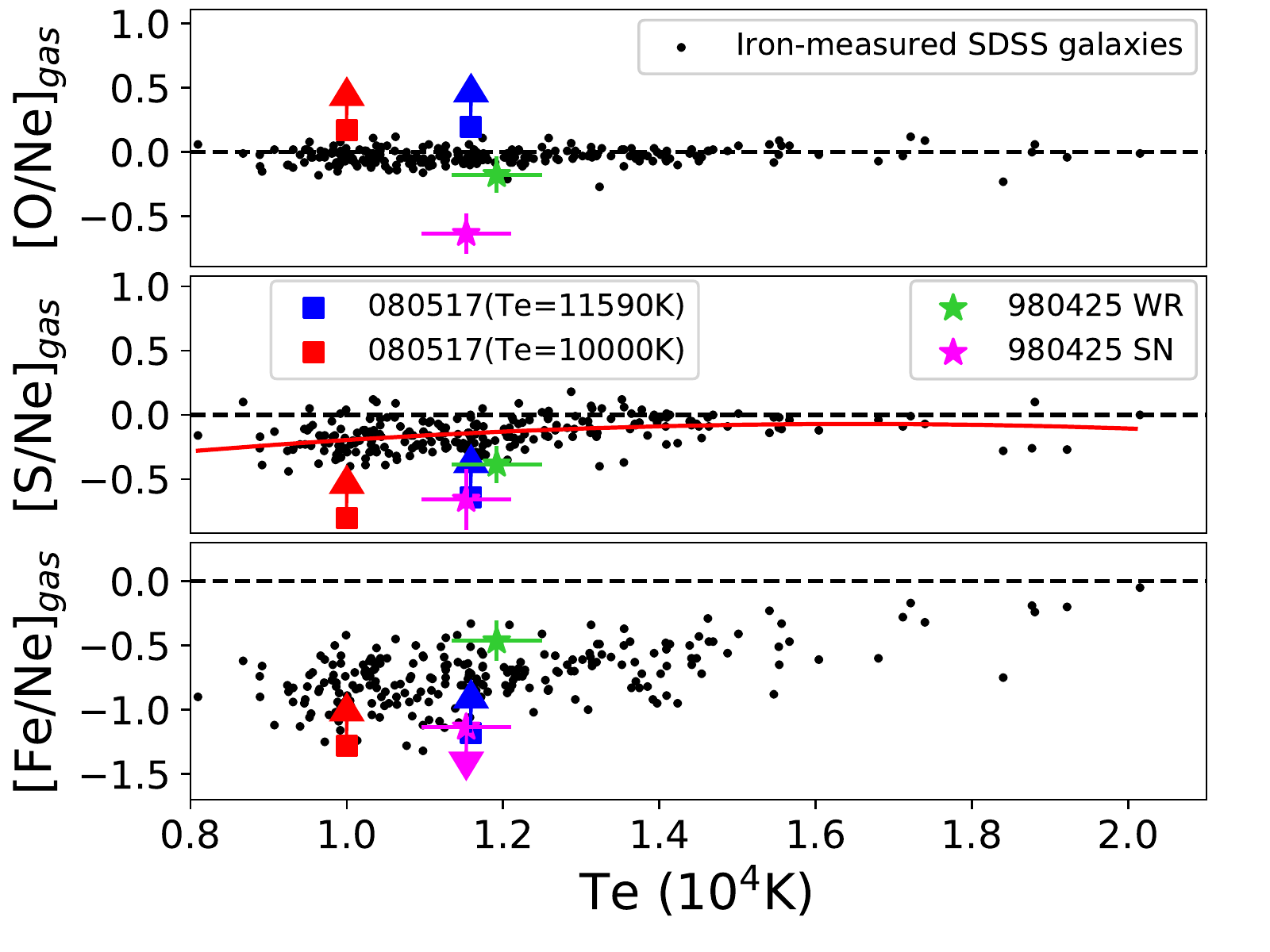}
    \caption{
    Gas-phase abundance ratios of the GRB 080517 and 980425 hosts (colored data points) as a function of electron temperature.
    Error bars contain observed errors of flux measurements and the statistical uncertainty of the abundance determination (0.1 dex) for each element.
    Gas-phase abundance ratios measured in the iron-measured SDSS galaxies are indicated by black dots.
    The solar value is show by horizontal dashed lines.
    The red line is the polynomial fitting of the [S/Ne]$_{gas}$--Te relation of the iron-measured SDSS galaxies. 
    }
    \label{fig12}
\end{figure}

Figure \ref{fig12} shows the [$X$/Ne]$_{gas}$ of the iron-measured SDSS galaxies and our GRB sample as a function of electron temperature. 
In the top panel of Figure \ref{fig12}, the iron-measured SDSS galaxies actually show the almost-constant value of [O/Ne]$_{gas}$=0.0 with the dispersion of the typical observational uncertainty of the abundance ratio, i.e., $\sim$0.14 dex, assuming the uncertainty of 0.1 dex \citep{Stasinska2005} for each element.
In general, oxygen is more similar to nonrefractory elements than sulfur or iron \citep[e.g.,][]{DeCia2016}. 
Therefore, the effect of the depletion of oxygen is very small, and the abundance ratio approaches the solar value.
The [O/Ne]$_{gas}$ of the iron-measured SDSS galaxies with lower electron temperatures is slightly lower than the solar value.
There may be a very weak dependency of the depletion factor of oxygen on electron temperature. 

This trend is more obvious in the middle panel of Figure \ref{fig12}. 
Sulfur is more easily depleted onto dust grains than oxygen. 
Thus, the [S/Ne]$_{gas}$ of the iron-measured SDSS galaxies shows a larger offset from the solar value compared with oxygen.
The dependency of [S/Ne]$_{gas}$ on electron temperature indicates that the depletion is more significant in the environment with the lower electron temperature. 
In the environment with lower the temperature, dust grains likely can survive and condense easily, in which abundant elements are depleted onto dust grains, i.e., the lower value of [S/Ne]$_{gas}$.
The higher electron temperature likely destroys more dust grains, which results in more abundant elements in gas, and the gas-phase abundance ratio approaches to the solar value. 
The significant variation of the depletion levels observed in galaxies is believed to be due to the dust condensation in the interstellar medium \citep[e.g.,][]{DeCia2016}.
Iron, an element with a high condensation temperature, below which 50\% or more of the element is removed from the gas phase at a pressure of 10$^{-4}$ atm, is more strongly depleted than an element with a low condensation temperature, such as oxygen \citep{Savage1996}.
This elemental-dependent depletion pattern is likely scaled by the temperature environment of the interstellar medium.
In fact, the gas-phase abundances in the cool diffuse cloud in the Milky Way are systematically more heavily depleted than those in the warm cloud \citep{Savage1996}.

The iron-to-neon abundance ratios (bottom panel in Figure \ref{fig12}) of the iron-measured SDSS galaxies indicate a further larger offset from the solar value.
This is expected from not only a stronger iron depletion than oxygen and sulfur but also the lesser contributions of iron enrichment by SNe Ia due to the young age (and/or top-heavy IMF) of the iron-measured SDSS galaxies. 

The observational biases of the iron-measured SDSS galaxies, i.e., low metallicity, possible young age, and top-heavy IMF, are basically related to the amount of $\alpha$ elements.
Supposing that the ratio of $\alpha$ elements cancels out such special characteristics, the ratio is determined by the depletion factor, which primarily depends on the temperature environment of the interstellar medium.
To estimate the depletion factor as a function of electron temperature, we performed a polynomial fitting for the [S/Ne]$_{gas}$--Te relation of the iron-measured SDSS galaxies (red line in the middle panel of Figure \ref{fig12}) rather than using the [O/Ne]$_{gas}$--Te relation. 
The offsets from the solar value in [O/Ne]$_{gas}$ are smaller than those in [S/Ne]$_{gas}$.
Therefore, [O/Ne]$_{gas}$ is easily affected by observational errors and less sensitive to estimating depletion factors than [S/Ne]$_{gas}$.
The best-fit function is
\begin{equation}
\label{delta_S}
[{\rm S/Ne}]_{gas}=\delta_{\rm S} = -0.872 + 0.971Te -0.294Te^{2}.
\end{equation}

\citet{DeCia2016} investigated the depletion factors of nine elements, including oxygen, sulfur, and iron, depending on the scaling factor of [Zn/Fe], based on 70 damped-Ly$\alpha$ absorbers (DLAs) toward quasars and depletion measurements in the Milky Way \citep{Jenkins2009}.
While the physical properties and chemical evolution of DLAs are probably different from those of the Milky Way, the depletion properties are linearly continuous in [Zn/Fe] all the way from the DLAs to the Galactic absorbers.
The depletion properties of the Small Magellanic Cloud and Large Magellanic Cloud \citep{Tchernyshyov2015} are also consistent with the depletion sequences as a function of [Zn/Fe] observed in DLAs and the Milky Way.
This suggests that the depletion properties can be investigated apart from the variety of individual characteristics of star-forming galaxies.
Therefore, we use their formulations to investigate the depletion properties of the iron-measured SDSS galaxies and GRB hosts.

\citet{DeCia2016} empirically characterized the trend of depletion factors of oxygen, sulfur, and iron with a factor of [Zn/Fe] as follows:
\begin{equation}
\delta_{\rm O}=-0.02-0.15{\rm [Zn/Fe]},
\end{equation}
\begin{equation}
\delta_{\rm S}=-0.04-0.28[{\rm Zn/Fe}],
\end{equation}
\begin{equation}
\delta_{\rm Fe}=-0.01-1.26[{\rm Zn/Fe}].
\end{equation}
By erasing the term of [Zn/Fe], we get the relation between the depletion factors of sulfur and iron, 
\begin{equation}
\label{delta_Fe}
\delta_{\rm Fe}=0.170+4.50\delta_{\rm S},
\end{equation}
and the relation between the depletion factors of sulfur and oxygen,
\begin{equation}
\label{delta_O}
\delta_{\rm O}=0.001 + 0.536\delta_{\rm S}
\end{equation}

Equations (\ref{delta_S}), (\ref{delta_Fe}), and (\ref{delta_O}) enable us to estimate the depletion factors of oxygen and iron from the depletion factor of sulfur. 
Based on the depletion factors of oxygen, sulfur, and iron of individual galaxies, we estimated the total abundances of these elements both in the iron-measured SDSS galaxies and in the GRB host sample.
The calculated gas-phase abundances, depletion factors, and total abundances of GRB 080517 and 980425 are summarized in table \ref{tab2}.

\begin{table*}
    \begin{center}
    \caption{
    Abundances and depletion factors of our sample.
    }
    \label{tab2}
     \begin{flushleft}
     \begin{tabular}{|l|l|l|l|l|c|}\hline \hline
     ID&080517&080517&980425 WR&980425 SN&Solar Value of 12+log($M$/H)\\ \relax
     $T_{e}$(O~{\sc iii})&11,590 K&10,000 K&11918 K&11,528 K & \\ \hline \relax
     [O/H]$_{gas}$    &-0.281$\pm$0.118$^{a}$& 0.042$\pm$0.119$^{a}$&-0.544$\pm$0.130$^{a}$&-0.372$\pm$0.139$^{a}$&8.69$^{e}$\\ \relax
     [Ne/H]$_{gas}$   &$<$-0.480$^{b}$ (3$\sigma$)&$<$-0.129$^{b}$ (3$\sigma$)&-0.369$\pm$0.137$^{a}$&0.260$\pm$0.174$^{a}$&7.93$^{e}$\\ \relax
     [S/H]$_{gas}$    &$>$-1.123$^{c}$&$>$-0.933$^{c}$&-0.753$\pm$0.132$^{a}$&-0.401$\pm$0.151$^{a}$&7.12$^{e}$\\ \relax
     [Fe/H]$_{gas}$   &-1.659$\pm$0.150$^{a}$&-1.410$\pm$0.149$^{a}$&-0.835$\pm$0.134$^{a}$&$<$-0.884$^{b}$ (3$\sigma$)&7.5$^{e}$\\ \relax
     [O/Ne]$_{gas}$   &$>$0.198$^{f}$&$>$0.170$^{f}$&-0.175$\pm$0.143$^{a}$&-0.631$\pm$0.159$^{a}$&-\\ \relax
     [S/Ne]$_{gas}$   &$>$-0.643$^{f}$&$>$-0.804$^{f}$&-0.383$\pm$0.145$^{a}$&-0.670$\pm$0.239$^{a}$&-\\ \relax
     [Fe/Ne]$_{gas}$  &$>$-1.179$^{f}$&$>$-1.281$^{f}$&-0.464$\pm$0.159$^{a}$&-1.146$\pm$0.193$^{a}$&-\\ \relax
     $\delta_{\rm O}$ &-0.074&-0.103&-0.069&-0.075&-\\ \relax
     $\delta_{\rm S}$ &-0.142&-0.195&-0.132&-0.143&-\\ \relax
     $\delta_{\rm Fe}$&-0.467&-0.707&-0.425&-0.475&-\\ \relax
     [O/H]$_{total}$  &-0.207$\pm$0.132$^{d}$& 0.145$\pm$0.133$^{d}$&-0.478$\pm$0.147$^{d}$&-0.296$\pm$0.152$^{d}$&8.69$^{e}$\\ \relax
     [S/H]$_{total}$  &$>$-0.981$^{c}$&$>$-0.738$^{c}$&-0.625$\pm$0.179$^{d}$&-0.262$\pm$0.186$^{d}$&7.12$^{e}$\\ \relax
     [Fe/H]$_{total}$ &-1.192$\pm$0.507$^{d}$&-0.703$\pm$0.514$^{d}$&-0.418$\pm$0.521$^{d}$&$<$-0.402$^{b}$ (3$\sigma$)&7.5$^{e}$\\ \relax
     [O/Ne]$_{total}$   &$>$0.273$^{f}$&$>$0.273$^{f}$&-0.105$\pm$0.154$^{d}$&-0.555$\pm$0.169$^{d}$&-\\ \relax
     [S/Ne]$_{total}$   &$>$-0.501$^{f}$&$>$-0.609$^{f}$&-0.248$\pm$0.181$^{d}$&-0.525$\pm$0.261$^{d}$&-\\ \relax
     [Fe/Ne]$_{total}$  &$>$-0.713$^{f}$&$>$-0.574$^{f}$&-0.026$\pm$0.516$^{d}$&$<$-0.663$^{b}$ (3$\sigma$)&-\\ \relax
     [O/Fe]$_{total}$ & 0.985$\pm$0.463$^{d}$& 0.847$\pm$0.471$^{d}$&-0.078$\pm$0.461$^{d}$&$>$0.102$^{b}$ (3$\sigma$)&-\\ \hline
     \end{tabular}\\
     $^{a}$Errors include observational errors of emission-line fluxes and the statistical uncertainty of the abundance calculation (0.1 dex).\\
     $^{b}$3$\sigma$ limit.\\
     $^{c}$Lower limit, since [S~{\sc ii}]$\lambda$6717 and $\lambda$6731 are detected but [S~{\sc iii}]$\lambda$6312 is not in the spectrum of the GRB 080517 host.\\
     $^{d}$Errors include the observational errors of the emission-line fluxes and statistical uncertainty of the abundance calculation (0.1 dex) and uncertainty of $\delta_{\rm S}$ (0.109 dex).\\
     $^{e}$\citet{Asplund2009}.\\
     $^{f}$Lower limit, since [Ne~{\sc iii}]$\lambda$3869 is not detected.\\
     \end{flushleft}
     \end{center}
\end{table*}

\subsection{Total iron abundance}
\label{total_iron}
\begin{figure}
	\includegraphics[width=8.5cm]{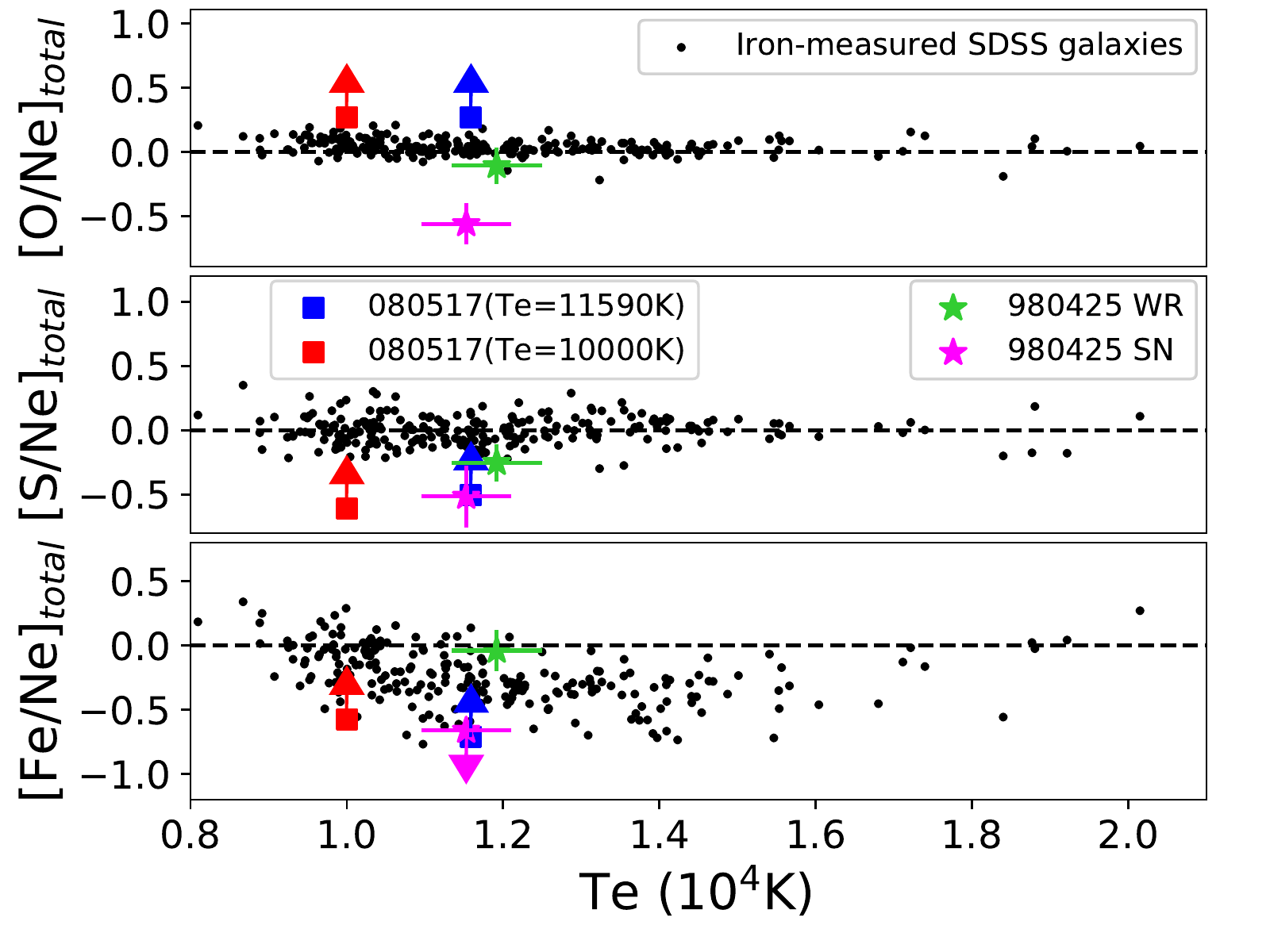}
    \caption{
    Same as Figure \ref{fig12} except for the total abundance ratios.
    }
    \label{fig13}
\end{figure}

As mentioned in section \ref{gas-phase_abundance}, the iron environment of GRBs should be examined by the total abundance.
Figure \ref{fig13} is the same as Figure \ref{fig12} but for the total abundance ratios.
The total abundance ratios of [O/Ne]$_{total}$ and [S/Ne]$_{total}$ of the iron-measured SDSS galaxies distribute around the solar value as expected. 
After the corrections for iron depletions, the [Fe/Ne]$_{total}$ of the iron-measured SDSS galaxies still deviates from the solar value.
The deviations likely reflect individual characteristics of different star-forming histories that result in different contributions of iron enrichment by SNe Ia and/or different IMFs.
Although only a lower limit is estimated for GRB 080517 due to the nondetection of [Ne~{\sc iii}]$\lambda$3869, the [Fe/Ne]$_{total}$ at the site of GRB 980425 is lower than the majority of the iron-measured SDSS galaxies, suggesting the low iron abundance nature of the GRB environment.

\begin{figure*}
    \begin{center}
	\includegraphics[width=17cm]{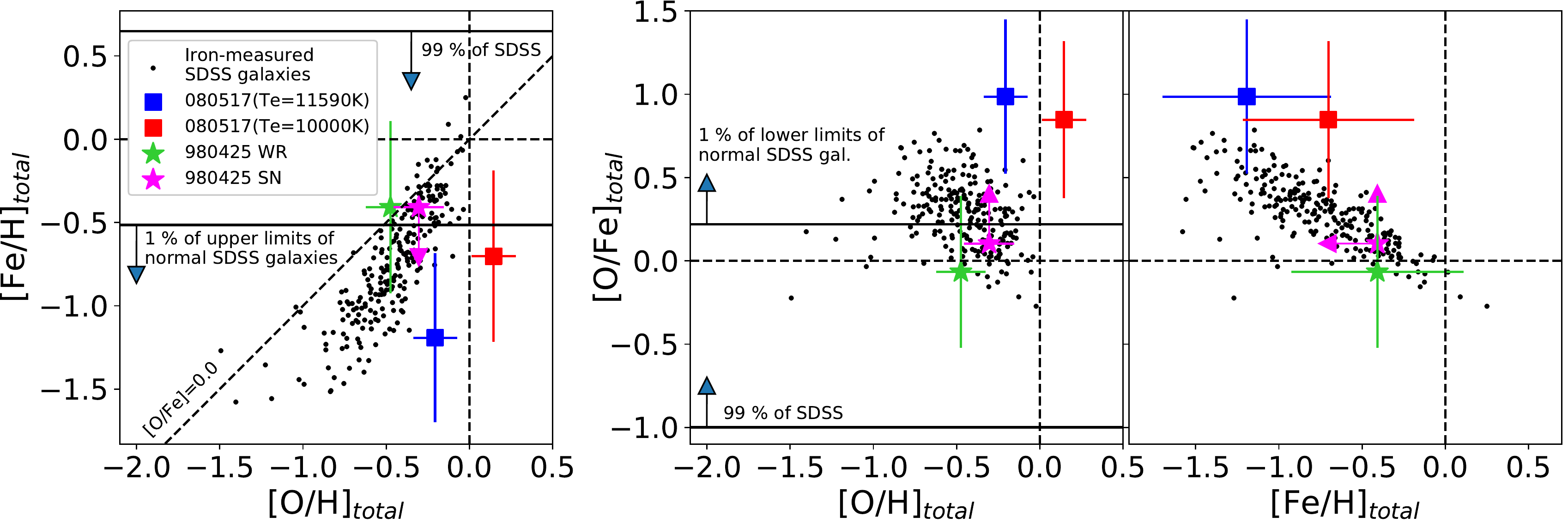}
    \caption{
    Same as Figure \ref{fig11} except for the total abundance ratios.
    Error bars contain the observational errors of flux measurements, statistical uncertainty of the abundance determination, and uncertainty of $\delta_{\rm S}$.
    The first and 99th percentiles of the upper/lower limits in normal star-forming galaxies selected from the SDSS DR7 are indicated by solid lines in the left/right panels.
    }
    \label{fig14}
    \end{center}
\end{figure*}

The low iron abundances of our GRB sample are also confirmed in Figure \ref{fig14}.
Figure \ref{fig14} is the same as Figure \ref{fig11} but for the total abundances, in which the error bars of the GRB sample contain the observational errors of the emission-line fluxes and statistical uncertainty of the abundance calculation (0.1 dex) and uncertainty of $\delta_{\rm S}$ (0.109 dex). 
In the left panel of Figure \ref{fig14}, the total oxygen and iron abundances of the iron-measured SDSS galaxies are well below the solar value.
The [Fe/H]$_{total}$ of the two GRB hosts is comparable to that of the iron-measured SDSS galaxies or still located at the lower part, even though the iron-measured SDSS galaxies are biased toward low metallicity.
Although the error bars are large, this trend is exactly same as that seen in Figure \ref{fig11}.
We also roughly estimated the upper limits of the [Fe/H]$_{total}$ of normal star-forming galaxies selected from the SDSS DR7.
We used the 3$\sigma$ uncertainty of the weak emission line in individual spectra to estimate the upper limit of [Fe~{\sc iii}]$\lambda$4658 flux and assumed an electron temperature of $T_{e}$(O~{\sc iii})=$1.0^{4}$ K.
The individual upper limits are derived in the same way as the iron-measured SDSS galaxies and GRB hosts.
The first and 99th percentiles of the upper limits in [Fe/H]$_{total}$ are shown by solid lines.
At least 1 \% of normal star-forming galaxies indicate [Fe/H]$_{total} \lesssim -0.5$, which is consistent with GRB host galaxies.
Note that the value of 1\% here is the lower limit, because the other 99\% might also show [Fe/H]$_{total} \lesssim -0.5$ due to the nondetections of the [Fe~{\sc iii}]$\lambda$4658 line in these galaxies.

In the right panel of Figure \ref{fig14}, the iron-measured SDSS galaxies with low [O/H]$_{total}$ (or low [Fe/H]$_{gas}$) reach at [O/Fe]$_{total}$ $\sim$0.5, which is roughly consistent with the convergence value of [O/Fe] determined by nucleosynthesis of SNe II without any contribution from SNe Ia. 
The [O/Fe]$_{total}$ of the iron-measured SDSS galaxies decreases with increasing [O/H]$_{total}$ (or increasing [Fe/H]$_{gas}$) down to around the solar value.
The GRB 080517 host shows quite high [O/Fe]$_{total}$, which is comparable to the highest values of [O/Fe]$_{total}$ of the iron-measured SDSS galaxies.
The lower limit of [O/Fe]$_{total}$ at the position of GRB 980425 is slightly higher than that of the WR region and the solar value.
The first and 99th percentiles of the lower limits of [O/Fe]$_{total}$ are estimated in normal star-forming galaxies (solid lines).
At least 1\% of normal star-forming galaxies selected from the SDSS DR7 indicate [O/Fe]$_{total} \gtrsim$ 0.22, which is consistent with GRB host galaxies.

\begin{figure}
	\includegraphics[width=8.5cm]{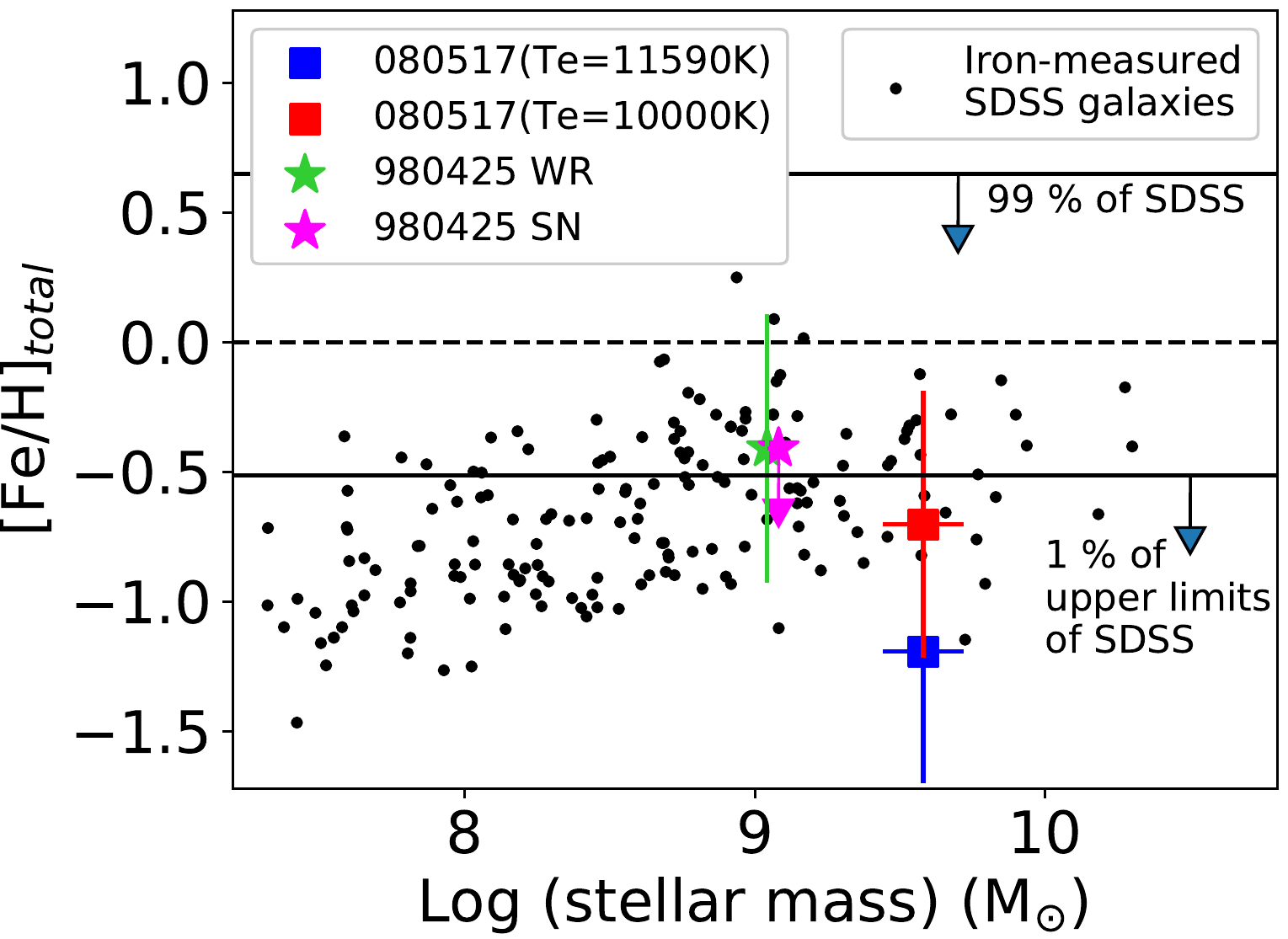}
    \caption{
    Stellar mass-iron abundance relation of the iron-measured SDSS galaxies.
    Symbols are the same as in Figure \ref{fig14}.
    }
    \label{fig15}
\end{figure}

Metallicity of star-forming galaxies is known to be primarily regulated by stellar mass \citep[e.g.,][]{Tremonti2004}.
There are correlations between stellar mass and oxygen abundance both in normal star-forming galaxies and in iron-measured SDSS galaxies in Figure \ref{fig3}.
The iron abundance is also expected to correlate with stellar mass, even though the iron-measured SDSS galaxies are biased toward low metallicity.
Therefore, the stellar mass-iron abundance relation is useful to clarify whether the low iron abundance of GRB hosts originates in the GRB mechanism or is just regulated by the stellar-mass effect.
Figure \ref{fig15} shows the stellar mass-iron abundance relation of the iron-measured SDSS galaxies.
The stellar mass correlates with iron abundance as expected.
The iron abundance of GRB 080517 is comparable to or below the stellar mass-iron abundance relation of the iron-measured SDSS galaxies, in contrast to the oxygen abundance, which is well above the stellar mass metallicity relation of the iron-measured SDSS galaxies, as shown in Figure \ref{fig3}.
The iron abundance of GRB 980425 is consistent with the stellar mass-iron abundance relation of the iron-measured SDSS galaxies.
Taking into account the possible bias of the iron-measured SDSS galaxies toward low metallicity, the stellar mass-metallicity relation based on iron abundance also supports the low iron abundance property of the GRB environment.

\section{discussion}
As suggested in section \ref{total_iron}, the GRB 080517 host and the site of GRB 980425 favor an environment with low iron abundance, even though the oxygen abundance of GRB 080517 is indeed comparable to the solar value.
The total iron abundances of these two GRB environments can be explained by the prediction of the single massive star scenario, i.e., [Fe/H]$\lesssim$ -1.0 \citep{Yoon2006}.

In general, GRB hosts are considered to be young star-forming galaxies \citep[e.g.,][]{Savaglio2009} and could be biased toward the top-heavy IMF if GRBs actually originate in single massive stars.
These two aspects of GRB hosts may yield the high value of [O/Fe] exceeding the solar value according to the following arguments.
Galaxies younger than $\sim 10^{9}$ yr probably have not yet experienced with chemical enrichment by SNe Ia which produce the majority of iron, but have experienced SNe II, which produce $\alpha$ elements such as oxygen, resulting in a high value of [O/Fe].
In addition, elemental yields of SN IIe show that [$\alpha$/Fe] increases with increasing progenitor mass \citep[e.g.,][]{Wyse1992,Woosley1995}.
Thus, top-heavy IMFs produce more abundant massive stars and higher [O/Fe].

In fact, the age of the GRB 080517 host indicated from the best-fit stellar component is $10^{9}$ yr (bottom panel of Figure \ref{fig4}).
\citet{Stanway2015} suggested the post-starburst stellar population with the age of 0.5$\times 10^{9}$ yr based on the spectral energy distribution from the ultraviolet through to the infrared. 
The host galaxy of GRB 080517 is likely young enough to lack the chemical enrichment by SNe Ia.
In the spectra of the GRB 980425 host, relatively older stellar populations of $\sim 10^{9.6}-10^{10.3}$ yr are underlying (bottom panels of Figure \ref{fig5} and \ref{fig6}) in both the WR region and GRB position.
Although there could be the contribution of the chemical enrichment by SNe Ia, GRB 980425 is still consistent with the iron-measured SDSS galaxies, i.e., galaxies biased toward low metallicity, in Figure \ref{fig14}.
This might suggest that the possible iron enrichment by SNe Ia in the GRB 980425 host is not sufficient to push it to the normal star-forming galaxy.

Given the Salpeter IMF, the prediction of nucleosynthesis models of SNe II yields [O/Fe]$_{total}$ $\sim$ 0.4.
This value can be as large as $\sim$ 0.7 if a top-heavy IMF slope is assumed \citep{Wyse1992}.
According to the evolutionary population synthesis models, the top-heavy IMF produces the larger fractional ratio of WR stars in young massive stars, because the proportion of WR progenitors among the new stellar generation is greater when the IMF is flatter \citep[e.g.,][]{Meynet1995}. 
Therefore, the clear evidence of WR stars indicates the top-heavy IMF in general.
The impact of the top-heavy IMF on [O/Fe]$_{total}$ in our sample is uncertain, because the S/N of the spectra of the GRB 080517 host and GRB 980425 site is not enough to identify the WR feature around the rest-frame 4868\AA.
The WR region in the GRB 980425 host does not show the high [O/Fe]$_{total}$ (Figure \ref{fig14}), in spite of the implied top-heavy IMF, which might suggest that the top-heavy IMF does not have a significant impact on [O/Fe]$_{total}$.

Since the error bars in Figure \ref{fig14} are large, we cannot rule out the possibility that the [O/Fe]$_{total}$ of GRB 080517 and explosion site of GRB 980425 are actually higher than the expectations from the young age and top-heavy IMF.
The oxygen-to-iron abundance ratio can be around [O/Fe]$_{total}\sim$1.0 within the error bars.
Such a high value of [O/Fe] is difficult to explain in terms of the young-age stellar population and top-heavy IMF.
One possible explanation is the fallback mechanism of the inner core materials of SNe II.
If the explosion energy is low, especially for massive stars, the ejected velocities of most of the $^{56}$Ni synthesized by explosive Si burning do not reach the escape velocity.  
Such $^{56}$Ni falls back onto the compact remnant, and only a small amount of $^{56}$Ni is mixed into the ejected materials.
Even for energetic explosions, fallback of a large amount of core materials (including $^{56}$Ni) can occur if the explosion is very aspherical to form a jet, and the rate of energy deposition in the jet is slow \citep{Tominaga2007}.  
In these SNe II, the small amount of ejected $^{56}$Ni decays into $^{56}$Fe via $^{56}$Co to power the light-curve tail, which is called faint SNe \citep[e.g.,][]{Nomoto2013}. 
In these SNe II with fallback, a large fraction of C, O, and other $\alpha$ elements in the outer layers is ejected.
As a result, yields of faint SNe II are expected to show high values of [$\alpha$/Fe] $\gtrsim$1.0.
This idea has been introduced to explain the abundance patterns of carbon-enhanced extremely metal-poor stars \citep[e.g.,][]{Tominaga2007}, hyper-metal-poor stars \citep[e.g.,][]{Iwamoto2005} and carbon-enhanced damped-Ly$\alpha$ systems \citep[e.g.,][]{Cooke2011,Kobayashi2011}.

The gamma-ray brightness of GRB 980425 is about 100 times fainter than that of other GRBs for which the associations with SNe are confirmed \citep{Hammer2006}.
However, the actual peak luminosity of the associated SN is unusually high compared with typical Type Ic SNe \citep{Galama1998}.
GRB 080517 is also categorized into the low-luminous GRBs \citep{Stanway2015}.
The SN component associated with the burst has not been detected.
Therefore, the fallback may be important in the environment of GRB 080517 rather than GRB 980425.
The observed sample is small, and the error bars of [O/Fe]$_{total}$ are large. 
Therefore, increasing the number of samples is quite important to investigate this hypothesis.

It would be worth mentioning the [Fe/H] implied in other GRB host galaxies.
The host galaxy of GRB 051117B indicates the highest oxygen abundance of GRB host galaxies \citep{Kruhler2015} at present.
The oxygen abundance is $12+\log({\rm O/H})=8.66$ based on the PP04N2 method \citep{Pettini2004}, which is calibrated to the Te method.
If the [O/Fe]$_{total}$ of the GRB 051117B host is 0.7-1.0 as seen in the 080517 host, the iron abundance of the 051117B host is estimated to be [Fe/H]=-1.0 to -0.7.
These values are still roughly consistent with the theoretical predictions of [Fe/H] $\lesssim$ -1.0.
Therefore, the high oxygen abundances found in some other GRB hosts \citep[e.g.,][]{Graham2013, Hashimoto2015, Kruhler2015} do not necessarily indicate high iron abundances or exclude the single massive star scenario.
On the other hand, if the [O/Fe] ratio is lower in the GRB 051117B host, then its [Fe/H] could be higher than the theoretical prediction of the single massive star scenario.

The iron abundance can also be estimated from the absorption system of GRB afterglow.
As in the case of average host metallicity based on emission-line diagnostics, it is not necessary for the absorption system to reflect the physical environment of the vicinity of GRBs.
The elemental abundances, including Zn, Si, S, and Fe, measured in GRB afterglows are actually lower than the solar value without any correction for the dust depletion \citep{Cucchiara2015}, although many of them are lower limits due to the poor spectral resolution.
\citet{Wiseman2017} investigated elemental abundances in 19 GRB afterglows based on high-resolution spectra with a correction for dust depletion.
In their high-resolution spectra, only one GRB shows an iron abundance comparable to the solar value after the depletion correction.
The observations of GRB afterglows basically support the low iron abundance environment of GRBs, except for a few high-metallicity cases \cite[e.g.,][]{Savaglio2012, Wiseman2017}.
We note that the abundance measurement in afterglow is severely biased toward optically bright events and is almost impossible in optically faint/dusty populations of GRB host galaxies, such as GRB 080517 in our sample.
In some cases, the emission line is still useful to measure metallicity even in dusty GRB host galaxies \citep[e.g.,][]{Hashimoto2015} and provides spatially resolved information that is essentially important to investigate the local environment around GRBs.
The systematic difference between emission- and absorption-based abundance measurements is also ambiguous.
Therefore, a complementary test from emission and absorption lines is necessary to reveal the physical environment of GRBs.

\section{Conclusions}
The widely accepted single massive star scenario predicts that GRBs occur in low-metallicity environments. 
In the scenario, a low iron abundance ([Fe/H]$\lesssim$-1.0) is required rather than oxygen for a massive star to launch the relativistic jet. 

However, oxygen has often been used to measure abundances of GRB host galaxies because its strong emission is easier to detect. 
Puzzlingly, the measured oxygen-based abundances of some GRB host galaxies have been higher than theoretically expected.

In this work, we measured iron abundance, which is more directly connected to the theory.
We obtained rest-frame optical spectra of two nearby GRB host galaxies (GRB 080517 and 980425).
We successfully detected weak [Fe~{\sc iii}]$\lambda$4658 emission lines in the spectra of the GRB 080517 host and WR region of GRB 980425 and constrained the upper limit of the iron abundance at the explosion site of GRB 980425.
The total iron abundances of the GRB 080517 host and explosion site of GRB 980425 are well below the solar value, even though the oxygen abundance of the GRB 080517 host is comparable to the solar value.
Although the error bars of iron abundances are large, the [Fe/H]$_{total}$ of our sample can be explained by the theoretical predictions ([Fe/H]$\lesssim$-1.0) of the single massive star scenario.
Relying only on oxygen abundance could mislead us on the origin of the GRBs.

The oxygen-to-iron ratio, [O/Fe]$_{total}$, of the GRB 080517 host is comparable to the highest value of the iron-measured SDSS galaxies.
The high [O/Fe] suggests that more oxygen has been produced in massive stars than iron from less massive stars. 
This situation can happen in (i) young galaxies where many type Ia SNe have not exploded yet or (ii) under a top-heavy IMF, where the fractions of massive stars are higher. 
The observed young stellar age ($\lesssim 10^{9}$ yr) of the GRB 080517 host supports scenario (i), though a conclusive argument on the importance of the scenario (ii) is difficult based on our small sample.
Alternatively, the fallback scenario of the SNe could also explain high [O/Fe]. 
In any case, iron is more directly related to the single massive star scenario of GRBs than oxygen. 
Our work is just the first step in this direction. 
In the future, it is important to investigate iron abundances of more GRB host galaxies. 

\acknowledgments
We thank the anonymous referee for many insightful comments.
We are very grateful to all of staff at the Subaru Telescope/VLT.
TG acknowledges the support by the Ministry of Science and Technology of Taiwan through grants NSC 103-2112-M-007-002-MY3 and 105-2112-M-007-003-MY3.
This work has been supported in part by the World Premier International Research Center Initiative (WPI Initiative), MEXT, Japan, and JSPS KAKENHI Grant Numbers JP26400222, JP16H02168, JP17K05382, and the Endowed Research Unit (Dark side of the universe) by Hamamatsu Photonics KK.

\bibliography{ApJ_GRB_iron} 

\begin{thebibliography}{}
\expandafter\ifx\csname natexlab\endcsname\relax\def\natexlab#1{#1}\fi
\providecommand{\url}[1]{\href{#1}{#1}}

\bibitem[{{Abazajian} {et~al.}(2009){Abazajian}, {Adelman-McCarthy},
  {Ag{\"u}eros}, {Allam}, {Allende Prieto}, {An}, {Anderson}, {Anderson},
  {Annis}, {Bahcall}, \& et~al.}]{Abazajian2009}
{Abazajian}, K.~N., {Adelman-McCarthy}, J.~K., {Ag{\"u}eros}, M.~A., {et~al.}
  2009, \apjs, 182, 543

\bibitem[{{Asplund} {et~al.}(2009){Asplund}, {Grevesse}, {Sauval}, \&
  {Scott}}]{Asplund2009}
{Asplund}, M., {Grevesse}, N., {Sauval}, A.~J., \& {Scott}, P. 2009, \araa, 47,
  481

\bibitem[{{Brinchmann} {et~al.}(2004){Brinchmann}, {Charlot}, {White},
  {Tremonti}, {Kauffmann}, {Heckman}, \& {Brinkmann}}]{Brinchmann2004}
{Brinchmann}, J., {Charlot}, S., {White}, S.~D.~M., {et~al.} 2004, \mnras, 351,
  1151

\bibitem[{{Calzetti} {et~al.}(2000){Calzetti}, {Armus}, {Bohlin}, {Kinney},
  {Koornneef}, \& {Storchi-Bergmann}}]{Calzetti2000}
{Calzetti}, D., {Armus}, L., {Bohlin}, R.~C., {et~al.} 2000, \apj, 533, 682

\bibitem[{{Cappellari}(2017)}]{Cappellari2017}
{Cappellari}, M. 2017, \mnras, 466, 798

\bibitem[{{Cappellari} \& {Emsellem}(2004)}]{Cappellari2004}
{Cappellari}, M., \& {Emsellem}, E. 2004, \pasp, 116, 138

\bibitem[{{Castro Cer{\'o}n} {et~al.}(2010){Castro Cer{\'o}n},
  {Micha{\l}owski}, {Hjorth}, {Malesani}, {Gorosabel}, {Watson}, {Fynbo}, \&
  {Morales Calder{\'o}n}}]{Castro2010}
{Castro Cer{\'o}n}, J.~M., {Micha{\l}owski}, M.~J., {Hjorth}, J., {et~al.}
  2010, \apj, 721, 1919

\bibitem[{{Christensen} {et~al.}(2008){Christensen}, {Vreeswijk}, {Sollerman},
  {Th{\"o}ne}, {Le Floc'h}, \& {Wiersema}}]{Christensen2008}
{Christensen}, L., {Vreeswijk}, P.~M., {Sollerman}, J., {et~al.} 2008, \aap,
  490, 45

\bibitem[{{Cid Fernandes} {et~al.}(2005){Cid Fernandes}, {Mateus}, {Sodr{\'e}},
  {Stasi{\'n}ska}, \& {Gomes}}]{Fernandes2005}
{Cid Fernandes}, R., {Mateus}, A., {Sodr{\'e}}, L., {Stasi{\'n}ska}, G., \&
  {Gomes}, J.~M. 2005, \mnras, 358, 363

\bibitem[{{Cid Fernandes} {et~al.}(2009){Cid Fernandes}, {Schoenell}, {Gomes},
  {Asari}, {Schlickmann}, {Mateus}, {Stasinska}, {Sodr{\'e}}, {Torres-Papaqui},
  \& {Seagal Collaboration}}]{Fernandes2009}
{Cid Fernandes}, R., {Schoenell}, W., {Gomes}, J.~M., {et~al.} 2009, in Revista
  Mexicana de Astronomia y Astrofisica Conference Series, Vol.~35, Revista
  Mexicana de Astronomia y Astrofisica Conference Series, 127--132

\bibitem[{{Cooke} {et~al.}(2011){Cooke}, {Pettini}, {Steidel}, {Rudie}, \&
  {Jorgenson}}]{Cooke2011}
{Cooke}, R., {Pettini}, M., {Steidel}, C.~C., {Rudie}, G.~C., \& {Jorgenson},
  R.~A. 2011, \mnras, 412, 1047

\bibitem[{{Cucchiara} {et~al.}(2015){Cucchiara}, {Fumagalli}, {Rafelski},
  {Kocevski}, {Prochaska}, {Cooke}, \& {Becker}}]{Cucchiara2015}
{Cucchiara}, A., {Fumagalli}, M., {Rafelski}, M., {et~al.} 2015, \apj, 804, 51

\bibitem[{{De Cia} {et~al.}(2016){De Cia}, {Ledoux}, {Mattsson}, {Petitjean},
  {Srianand}, {Gavignaud}, \& {Jenkins}}]{DeCia2016}
{De Cia}, A., {Ledoux}, C., {Mattsson}, L., {et~al.} 2016, \aap, 596, A97

\bibitem[{{Fryer} {et~al.}(1999){Fryer}, {Woosley}, \& {Hartmann}}]{Fryer1999}
{Fryer}, C.~L., {Woosley}, S.~E., \& {Hartmann}, D.~H. 1999, \apj, 526, 152

\bibitem[{{Galama} {et~al.}(1998){Galama}, {Vreeswijk}, {van Paradijs},
  {Kouveliotou}, {Augusteijn}, {B{\"o}hnhardt}, {Brewer}, {Doublier},
  {Gonzalez}, {Leibundgut}, {Lidman}, {Hainaut}, {Patat}, {Heise}, {in't Zand},
  {Hurley}, {Groot}, {Strom}, {Mazzali}, {Iwamoto}, {Nomoto}, {Umeda},
  {Nakamura}, {Young}, {Suzuki}, {Shigeyama}, {Koshut}, {Kippen}, {Robinson},
  {de Wildt}, {Wijers}, {Tanvir}, {Greiner}, {Pian}, {Palazzi}, {Frontera},
  {Masetti}, {Nicastro}, {Feroci}, {Costa}, {Piro}, {Peterson}, {Tinney},
  {Boyle}, {Cannon}, {Stathakis}, {Sadler}, {Begam}, \& {Ianna}}]{Galama1998}
{Galama}, T.~J., {Vreeswijk}, P.~M., {van Paradijs}, J., {et~al.} 1998, \nat,
  395, 670

\bibitem[{{Graham} \& {Fruchter}(2013)}]{Graham2013}
{Graham}, J.~F., \& {Fruchter}, A.~S. 2013, \apj, 774, 119

\bibitem[{{Hammer} {et~al.}(2006){Hammer}, {Flores}, {Schaerer},
  {Dessauges-Zavadsky}, {Le Floc'h}, \& {Puech}}]{Hammer2006}
{Hammer}, F., {Flores}, H., {Schaerer}, D., {et~al.} 2006, \aap, 454, 103

\bibitem[{{Hashimoto} {et~al.}(2018){Hashimoto}, {Goto}, \&
  {Momose}}]{Hashimoto2018}
{Hashimoto}, T., {Goto}, T., \& {Momose}, R. 2018, \mnras, 475, 4424

\bibitem[{{Hashimoto} {et~al.}(2015){Hashimoto}, {Perley}, {Ohta}, {Aoki},
  {Tanaka}, {Niino}, {Yabe}, \& {Kawai}}]{Hashimoto2015}
{Hashimoto}, T., {Perley}, D.~A., {Ohta}, K., {et~al.} 2015, \apj, 806, 250

\bibitem[{{Hashimoto} {et~al.}(2010){Hashimoto}, {Ohta}, {Aoki}, {Tanaka},
  {Yabe}, {Kawai}, {Aoki}, {Furusawa}, {Hattori}, {Iye}, {Kawabata},
  {Kobayashi}, {Komiyama}, {Kosugi}, {Minowa}, {Mizumoto}, {Niino}, {Nomoto},
  {Noumaru}, {Ogasawara}, {Pyo}, {Sakamoto}, {Sekiguchi}, {Shirasaki},
  {Suzuki}, {Tajitsu}, {Takata}, {Tamagawa}, {Terada}, {Totani}, {Watanabe},
  {Yamada}, \& {Yoshida}}]{Hashimoto2010}
{Hashimoto}, T., {Ohta}, K., {Aoki}, K., {et~al.} 2010, \apj, 719, 378

\bibitem[{{Iwamoto} {et~al.}(2005){Iwamoto}, {Umeda}, {Tominaga}, {Nomoto}, \&
  {Maeda}}]{Iwamoto2005}
{Iwamoto}, N., {Umeda}, H., {Tominaga}, N., {Nomoto}, K., \& {Maeda}, K. 2005,
  Science, 309, 451

\bibitem[{{Izotov} {et~al.}(2006){Izotov}, {Stasi{\'n}ska}, {Meynet}, {Guseva},
  \& {Thuan}}]{Izotov2006}
{Izotov}, Y.~I., {Stasi{\'n}ska}, G., {Meynet}, G., {Guseva}, N.~G., \&
  {Thuan}, T.~X. 2006, \aap, 448, 955

\bibitem[{{Jenkins}(2009)}]{Jenkins2009}
{Jenkins}, E.~B. 2009, \apj, 700, 1299

\bibitem[{{Kashikawa} {et~al.}(2002){Kashikawa}, {Aoki}, {Asai}, {Ebizuka},
  {Inata}, {Iye}, {Kawabata}, {Kosugi}, {Ohyama}, {Okita}, {Ozawa}, {Saito},
  {Sasaki}, {Sekiguchi}, {Shimizu}, {Taguchi}, {Takata}, {Yadoumaru}, \&
  {Yoshida}}]{Kashikawa2002}
{Kashikawa}, N., {Aoki}, K., {Asai}, R., {et~al.} 2002, \pasj, 54, 819

\bibitem[{{Kauffmann} {et~al.}(2003){Kauffmann}, {Heckman}, {Tremonti},
  {Brinchmann}, {Charlot}, {White}, {Ridgway}, {Brinkmann}, {Fukugita}, {Hall},
  {Ivezi{\'c}}, {Richards}, \& {Schneider}}]{Kauffmann2003}
{Kauffmann}, G., {Heckman}, T.~M., {Tremonti}, C., {et~al.} 2003, \mnras, 346,
  1055

\bibitem[{{Kobayashi} {et~al.}(2011){Kobayashi}, {Tominaga}, \&
  {Nomoto}}]{Kobayashi2011}
{Kobayashi}, C., {Tominaga}, N., \& {Nomoto}, K. 2011, \apjl, 730, L14

\bibitem[{{Kouveliotou} {et~al.}(1993){Kouveliotou}, {Meegan}, {Fishman},
  {Bhat}, {Briggs}, {Koshut}, {Paciesas}, \& {Pendleton}}]{Kouveliotou1993}
{Kouveliotou}, C., {Meegan}, C.~A., {Fishman}, G.~J., {et~al.} 1993, \apjl,
  413, L101

\bibitem[{{Kr{\"u}hler} {et~al.}(2017){Kr{\"u}hler}, {Kuncarayakti}, {Schady},
  {Anderson}, {Galbany}, \& {Gensior}}]{Kruhler2017}
{Kr{\"u}hler}, T., {Kuncarayakti}, H., {Schady}, P., {et~al.} 2017, \aap, 602,
  A85

\bibitem[{{Kr{\"u}hler} {et~al.}(2015){Kr{\"u}hler}, {Malesani}, {Fynbo},
  {Hartoog}, {Hjorth}, {Jakobsson}, {Perley}, {Rossi}, {Schady}, {Schulze},
  {Tanvir}, {Vergani}, {Wiersema}, {Afonso}, {Bolmer}, {Cano}, {Covino},
  {D'Elia}, {de Ugarte Postigo}, {Filgas}, {Friis}, {Graham}, {Greiner},
  {Goldoni}, {Gomboc}, {Hammer}, {Japelj}, {Kann}, {Kaper}, {Klose}, {Levan},
  {Leloudas}, {Milvang-Jensen}, {Nicuesa Guelbenzu}, {Palazzi}, {Pian},
  {Piranomonte}, {S{\'a}nchez-Ram{\'{\i}}rez}, {Savaglio}, {Selsing},
  {Tagliaferri}, {Vreeswijk}, {Watson}, \& {Xu}}]{Kruhler2015}
{Kr{\"u}hler}, T., {Malesani}, D., {Fynbo}, J.~P.~U., {et~al.} 2015, \aap, 581,
  A125

\bibitem[{{Levesque} {et~al.}(2010{\natexlab{a}}){Levesque}, {Berger},
  {Kewley}, \& {Bagley}}]{Levesque2010a}
{Levesque}, E.~M., {Berger}, E., {Kewley}, L.~J., \& {Bagley}, M.~M.
  2010{\natexlab{a}}, \aj, 139, 694

\bibitem[{{Levesque} {et~al.}(2010{\natexlab{b}}){Levesque}, {Kewley},
  {Berger}, \& {Zahid}}]{Levesque2010b}
{Levesque}, E.~M., {Kewley}, L.~J., {Berger}, E., \& {Zahid}, H.~J.
  2010{\natexlab{b}}, \aj, 140, 1557

\bibitem[{{Mannucci} {et~al.}(2010){Mannucci}, {Cresci}, {Maiolino}, {Marconi},
  \& {Gnerucci}}]{Mannucci2010}
{Mannucci}, F., {Cresci}, G., {Maiolino}, R., {Marconi}, A., \& {Gnerucci}, A.
  2010, \mnras, 408, 2115

\bibitem[{{McWilliam}(1997)}]{McWilliam1997}
{McWilliam}, A. 1997, \araa, 35, 503

\bibitem[{{Meynet}(1995)}]{Meynet1995}
{Meynet}, G. 1995, \aap, 298, 767

\bibitem[{{Modjaz} {et~al.}(2008){Modjaz}, {Kewley}, {Kirshner}, {Stanek},
  {Challis}, {Garnavich}, {Greene}, {Kelly}, \& {Prieto}}]{Modjaz2008}
{Modjaz}, M., {Kewley}, L., {Kirshner}, R.~P., {et~al.} 2008, \aj, 135, 1136

\bibitem[{{Niino} {et~al.}(2015){Niino}, {Nagamine}, \& {Zhang}}]{Niino2015}
{Niino}, Y., {Nagamine}, K., \& {Zhang}, B. 2015, \mnras, 449, 2706

\bibitem[{{Nomoto} {et~al.}(1995){Nomoto}, {Iwamoto}, \& {Suzuki}}]{Nomoto1995}
{Nomoto}, K., {Iwamoto}, K., \& {Suzuki}, T. 1995, \physrep, 256, 173

\bibitem[{{Nomoto} {et~al.}(2013){Nomoto}, {Kobayashi}, \&
  {Tominaga}}]{Nomoto2013}
{Nomoto}, K., {Kobayashi}, C., \& {Tominaga}, N. 2013, \araa, 51, 457

\bibitem[{{Oliva} {et~al.}(2001){Oliva}, {Marconi}, {Maiolino}, {Testi},
  {Mannucci}, {Ghinassi}, {Licandro}, {Origlia}, {Baffa}, {Checcucci},
  {Comoretto}, {Gavryussev}, {Gennari}, {Giani}, {Hunt}, {Lisi}, {Lorenzetti},
  {Marcucci}, {Miglietta}, {Sozzi}, {Stefanini}, \& {Vitali}}]{Oliva2001}
{Oliva}, E., {Marconi}, A., {Maiolino}, R., {et~al.} 2001, \aap, 369, L5

\bibitem[{{Onodera} {et~al.}(2012){Onodera}, {Renzini}, {Carollo},
  {Cappellari}, {Mancini}, {Strazzullo}, {Daddi}, {Arimoto}, {Gobat}, {Yamada},
  {McCracken}, {Ilbert}, {Capak}, {Cimatti}, {Giavalisco}, {Koekemoer}, {Kong},
  {Lilly}, {Motohara}, {Ohta}, {Sanders}, {Scoville}, {Tamura}, \&
  {Taniguchi}}]{Onodera2012}
{Onodera}, M., {Renzini}, A., {Carollo}, M., {et~al.} 2012, \apj, 755, 26

\bibitem[{{Osterbrock}(1989)}]{Osterbrock1989}
{Osterbrock}, D.~E. 1989, {Astrophysics of gaseous nebulae and active galactic
  nuclei}

\bibitem[{{Perley} {et~al.}(2016{\natexlab{a}}){Perley}, {Kr{\"u}hler},
  {Schulze}, {de Ugarte Postigo}, {Hjorth}, {Berger}, {Cenko}, {Chary},
  {Cucchiara}, {Ellis}, {Fong}, {Fynbo}, {Gorosabel}, {Greiner}, {Jakobsson},
  {Kim}, {Laskar}, {Levan}, {Micha{\l}owski}, {Milvang-Jensen}, {Tanvir},
  {Th{\"o}ne}, \& {Wiersema}}]{Perley2016a}
{Perley}, D.~A., {Kr{\"u}hler}, T., {Schulze}, S., {et~al.} 2016{\natexlab{a}},
  \apj, 817, 7

\bibitem[{{Perley} {et~al.}(2016{\natexlab{b}}){Perley}, {Tanvir}, {Hjorth},
  {Laskar}, {Berger}, {Chary}, {de Ugarte Postigo}, {Fynbo}, {Kr{\"u}hler},
  {Levan}, {Micha{\l}owski}, \& {Schulze}}]{Perley2016b}
{Perley}, D.~A., {Tanvir}, N.~R., {Hjorth}, J., {et~al.} 2016{\natexlab{b}},
  \apj, 817, 8

\bibitem[{{Pettini} \& {Pagel}(2004)}]{Pettini2004}
{Pettini}, M., \& {Pagel}, B.~E.~J. 2004, \mnras, 348, L59

\bibitem[{{Prochaska} {et~al.}(2007){Prochaska}, {Chen}, {Dessauges-Zavadsky},
  \& {Bloom}}]{Prochaska2007}
{Prochaska}, J.~X., {Chen}, H.-W., {Dessauges-Zavadsky}, M., \& {Bloom}, J.~S.
  2007, \apj, 666, 267

\bibitem[{{Proxauf} {et~al.}(2014){Proxauf}, {{\"O}ttl}, \&
  {Kimeswenger}}]{Proxauf2014}
{Proxauf}, B., {{\"O}ttl}, S., \& {Kimeswenger}, S. 2014, \aap, 561, A10

\bibitem[{{Rodr{\'{\i}}guez}(2002)}]{Rodriguez2002}
{Rodr{\'{\i}}guez}, M. 2002, \aap, 389, 556

\bibitem[{{Salim} {et~al.}(2007){Salim}, {Rich}, {Charlot}, {Brinchmann},
  {Johnson}, {Schiminovich}, {Seibert}, {Mallery}, {Heckman}, {Forster},
  {Friedman}, {Martin}, {Morrissey}, {Neff}, {Small}, {Wyder}, {Bianchi},
  {Donas}, {Lee}, {Madore}, {Milliard}, {Szalay}, {Welsh}, \& {Yi}}]{Salim2007}
{Salim}, S., {Rich}, R.~M., {Charlot}, S., {et~al.} 2007, \apjs, 173, 267

\bibitem[{{Savage} \& {Sembach}(1996)}]{Savage1996}
{Savage}, B.~D., \& {Sembach}, K.~R. 1996, \araa, 34, 279

\bibitem[{{Savaglio} {et~al.}(2009){Savaglio}, {Glazebrook}, \& {Le
  Borgne}}]{Savaglio2009}
{Savaglio}, S., {Glazebrook}, K., \& {Le Borgne}, D. 2009, \apj, 691, 182

\bibitem[{{Savaglio} {et~al.}(2012){Savaglio}, {Rau}, {Greiner}, {Kr{\"u}hler},
  {McBreen}, {Hartmann}, {Updike}, {Filgas}, {Klose}, {Afonso}, {Clemens},
  {K{\"u}pc{\"u} Yolda{\c s}}, {Olivares E.}, {Sudilovsky}, \&
  {Szokoly}}]{Savaglio2012}
{Savaglio}, S., {Rau}, A., {Greiner}, J., {et~al.} 2012, \mnras, 420, 627

\bibitem[{{Seaton}(1979)}]{Seaton1979}
{Seaton}, M.~J. 1979, \mnras, 187, 73P

\bibitem[{{Stanek} {et~al.}(2006){Stanek}, {Gnedin}, {Beacom}, {Gould},
  {Johnson}, {Kollmeier}, {Modjaz}, {Pinsonneault}, {Pogge}, \&
  {Weinberg}}]{Stanek2006}
{Stanek}, K.~Z., {Gnedin}, O.~Y., {Beacom}, J.~F., {et~al.} 2006, \actaa, 56,
  333

\bibitem[{{Stanway} {et~al.}(2015){Stanway}, {Levan}, {Tanvir}, {Wiersema},
  {van der Horst}, {Mundell}, \& {Guidorzi}}]{Stanway2015}
{Stanway}, E.~R., {Levan}, A.~J., {Tanvir}, N., {et~al.} 2015, \mnras, 446,
  3911

\bibitem[{{Stasi{\'n}ska}(2005)}]{Stasinska2005}
{Stasi{\'n}ska}, G. 2005, \aap, 434, 507

\bibitem[{{Steidel} {et~al.}(2016){Steidel}, {Strom}, {Pettini}, {Rudie},
  {Reddy}, \& {Trainor}}]{Steidel2016}
{Steidel}, C.~C., {Strom}, A.~L., {Pettini}, M., {et~al.} 2016, \apj, 826, 159

\bibitem[{{Tchernyshyov} {et~al.}(2015){Tchernyshyov}, {Meixner}, {Seale},
  {Fox}, {Friedman}, {Dwek}, \& {Galliano}}]{Tchernyshyov2015}
{Tchernyshyov}, K., {Meixner}, M., {Seale}, J., {et~al.} 2015, \apj, 811, 78

\bibitem[{{Tinsley}(1979)}]{Tinsley1979}
{Tinsley}, B.~M. 1979, \apj, 229, 1046

\bibitem[{{Tominaga} {et~al.}(2007){Tominaga}, {Maeda}, {Umeda}, {Nomoto},
  {Tanaka}, {Iwamoto}, {Suzuki}, \& {Mazzali}}]{Tominaga2007}
{Tominaga}, N., {Maeda}, K., {Umeda}, H., {et~al.} 2007, \apjl, 657, L77

\bibitem[{{Tremonti} {et~al.}(2004){Tremonti}, {Heckman}, {Kauffmann},
  {Brinchmann}, {Charlot}, {White}, {Seibert}, {Peng}, {Schlegel}, {Uomoto},
  {Fukugita}, \& {Brinkmann}}]{Tremonti2004}
{Tremonti}, C.~A., {Heckman}, T.~M., {Kauffmann}, G., {et~al.} 2004, \apj, 613,
  898

\bibitem[{{Vazdekis} {et~al.}(2010){Vazdekis}, {S{\'a}nchez-Bl{\'a}zquez},
  {Falc{\'o}n-Barroso}, {Cenarro}, {Beasley}, {Cardiel}, {Gorgas}, \&
  {Peletier}}]{Vazdekis2010}
{Vazdekis}, A., {S{\'a}nchez-Bl{\'a}zquez}, P., {Falc{\'o}n-Barroso}, J.,
  {et~al.} 2010, \mnras, 404, 1639

\bibitem[{{Vink} \& {de Koter}(2005)}]{Vink2005}
{Vink}, J.~S., \& {de Koter}, A. 2005, \aap, 442, 587

\bibitem[{{Wiseman} {et~al.}(2017){Wiseman}, {Schady}, {Bolmer}, {Kr{\"u}hler},
  {Yates}, {Greiner}, \& {Fynbo}}]{Wiseman2017}
{Wiseman}, P., {Schady}, P., {Bolmer}, J., {et~al.} 2017, \aap, 599, A24

\bibitem[{{Woosley} \& {Heger}(2006)}]{Woosley2006}
{Woosley}, S.~E., \& {Heger}, A. 2006, \apj, 637, 914

\bibitem[{{Woosley} \& {Weaver}(1995)}]{Woosley1995}
{Woosley}, S.~E., \& {Weaver}, T.~A. 1995, \apjs, 101, 181

\bibitem[{{Wyse} \& {Gilmore}(1992)}]{Wyse1992}
{Wyse}, R.~F.~G., \& {Gilmore}, G. 1992, \aj, 104, 144

\bibitem[{{Yoon} {et~al.}(2006){Yoon}, {Langer}, \& {Norman}}]{Yoon2006}
{Yoon}, S.-C., {Langer}, N., \& {Norman}, C. 2006, \aap, 460, 199

\end{thebibliography}



\end{document}